\documentclass[epsfig,12pt]{article}
\usepackage{epsfig,amssymb,euscript,bbold}
\newcommand{\be}{\begin{equation}}
\newcommand{\ee}{\end{equation}}
\newcommand{\bea}{\begin{eqnarray}}
\newcommand{\eea}{\end{eqnarray}}
\newcommand{\ba}{\begin{array}}
\newcommand{\p}[1]{(\ref{#1})}
\newcommand{\ea}{\end{array}}

\def\bbox{{\,\lower0.9pt\vbox{\hrule \hbox{\vrule height 0.2 cm
\hskip 0.2 cm \vrule height 0.2 cm}\hrule}\,}}
\newcommand{\dsl}{\pa \kern-0.5em /}

\newcommand{\nn}{\nonumber \\}

\newcommand{\basismu}{\{1,\Gamma_{\a_1},\Gamma_{\a_1\a_2},\Gamma_{\a_1\a_2\a_3},\Gamma_{\a_1\a_2\a_3\a_4},\Gamma_{\a_1\a_2\a_3\a_4\a_5}\}}

\def\CL{{\cal L}}                       

\def\ds{\raise.15ex\hbox{/}\kern-.57em\partial}
\def\Ds{\,\raise.15ex\hbox{/}\mkern-13.5mu D}

\renewcommand{\a}{\alpha}
\renewcommand{\b}{\beta}

\newcommand{\e}{\epsilon}
\newcommand{\K}{{{K}}}

\font\mybb=msbm10 at 10pt
\def\bb#1{\hbox{\mybb#1}}

\def\bR {\bb{R}}
\def\bE {\bb{E}}

\def\a{\alpha}\def\b{\beta}\def\l{\lambda}
\def\s{\sigma}

\begin{document}

\makeatletter
\renewcommand{\theequation}{\thesection.\arabic{equation}}
\@addtoreset{equation}{section} \makeatother

\baselineskip 18pt


\begin{titlepage}
\vfill
\begin{flushright}
QMUL-PH-02-22\\
hep-th/0212008\\
\end{flushright}

\vfill

\begin{center}
\baselineskip=16pt {\Large\bf The Geometry of D=11 Killing
Spinors}
\vskip 10.mm {Jerome P. Gauntlett$^{1}$ and Stathis Pakis$^{2}$\\}
\vskip 1cm
{\small\it
Department of Physics\\
Queen Mary, University of London\\
Mile End Rd, London E1 4NS, U.K.\\} \vskip 0.5cm \vskip 0.5cm
\vspace{6pt}
\end{center}
\vfill
\par
\begin{center}
{\bf ABSTRACT}
\end{center}
\begin{quote}
We propose a way to classify the local form of all bosonic
supersymmetric configurations of D=11 supergravity, using the
differential forms that can be constructed as bi-linears from the
Killing spinors. We show that the most general bosonic geometries
either have a privileged $SU(5)$ or a $(Spin(7)\ltimes\bR^8)\times
\bR$ structure, depending on whether the Killing vector
constructed from the Killing spinor is timelike or null,
respectively. In the time-like case we derive the general local
form of the geometry and show that it is almost completely
determined by a certain $SU(5)$ structure on the ten-dimensional
space orthogonal to the orbits of the Killing vector. We also
deduce what further conditions must be imposed in order that the
equations of motion are satisfied. We illustrate the formalism
with some known solutions and also present some new solutions
including a rotating generalisation of the resolved membrane
solutions and generalisations of the recently constructed D=11
G\"odel solution. We also prove some general vanishing theorems
for compactifications with flux.

\vfill \vskip 5mm \hrule width 5.cm \vskip 5mm {\small
\noindent $^1$ E-mail: j.p.gauntlett@qmul.ac.uk \\
\noindent $^2$ E-mail: s.pakis@qmul.ac.uk \\
}
\end{quote}
\end{titlepage}
\setcounter{equation}{0}

\section{Introduction}
Supersymmetric solutions of supergravity theories have played a
prominent role in many developments in string theory and it would
be useful to have a systematic classification of all such
solutions. When the fluxes are all set to zero we know that the
supersymmetric geometries must admit covariantly constant spinors
and hence must admit metrics with special holonomy. In the
Riemannian case the possible special holonomy groups that can
arise are completely classified. {}For a relevant discussion
on the Lorentzian case, see \cite{bryant,Figueroa-O'Farrill:1999tx}.

There are many results in the literature concerning special cases
when the fluxes are non-vanishing but they are typically based on
special ansatz\"e\footnote{An exception is the recent
classification of maximally supersymmetric solutions of D=10,11
supergravity \cite{josegeorge}.} and a global
picture has been lacking. Here we shall propose a way to classify
the local forms of all supersymmetric solutions of D=11
supergravity, independent of ansatz, building on the work of
\cite{friv,Gauntlett:2001ur,iv,frivtwo,Gauntlett:2002sc,Gauntlett:2002nw}
using the $G$-structures defined by Killing spinors\footnote{{}For
other work relating $G$-structures to supergravity solutions with
non-vanishing fluxes, see \cite{Gurrieri:2002wz,Cardoso:2002hd}.}.
Moreover, it is clear how to extend the ideas to any supergravity
theory. Indeed a complete analysis for $D=5$ minimal supergravity
has already been carried out in \cite{Gauntlett:2002nw} (for
earlier work on the simpler case of $N=2$ supergravity in $D=4$,
using techniques specific to $D=4$, see \cite{tod}).

We should mention at the outset, that the classification we are
advocating would still leave the very challenging task of
determining all of the explicit supersymmetric solutions that can
arise within the classes we will discuss. For example, in the
special case when the flux is zero, this corresponds to
explicitly classifying all special holonomy manifolds, which seems to 
require
fundamentally new mathematical ideas in order to make
progress.

We start in section 2 by deriving a number of necessary conditions
for a bosonic geometry, consisting of a metric and a four-form, to
admit Killing spinors. We first construct differential forms of
rank 0,...,5 from bi-linears of the Killing spinors. {}Fierz
identities lead to a number of algebraic conditions that these
forms must satisfy, while the Killing spinor equation gives a
number of differential constraints. For example, the vector fields
dual to the one-forms $K$ constructed from the Killing spinors are
always Killing. When one of the $K$ is timelike, we show that some
of the differential conditions are those of generalised
calibrations \cite{Gutowski:1999iu,gpt,blw} for membranes and also
for fivebranes, with a small extension for the latter case. The
same differential conditions hold when $K$ is null or, when there
is more than one Killing spinor, spacelike, which suggests an
interesting extension of the notion of generalised calibration.

In section 3, we argue that $G$-structures are very useful for
interpreting and organising the results of section 2. We begin by
recalling the notion of $G$-structures and their classification
and then discuss how they can provide the basis for a
classification of all supersymmetric solutions. One important
result is that any supersymmetric solution, i.e. preserving at
least one Killing spinor (1/32 supersymmetry), will either have a
privileged local $SU(5)$ or an $(Spin(7)\ltimes\bR^8)\times \bR$
structure that is constructed from the Killing spinor. The two
cases are distinguished by whether the Killing vector is time-like
or null, respectively.

To be more precise, the Killing vector is either null everywhere
or it isn't. The former ``null''-case admits a
globally\footnote{We will ignore issues to do with precisely what
kinds of singularities we want to allow in physically relevant
solutions.} defined $(Spin(7)\ltimes\bR^8)\times \bR$ structure.
In the latter ``time-like''-case there is at least a point and
hence a neighbourhood where $K$ is timelike. Since the
neighbourhood is topologically trivial, the frame bundle can
always be trivialised. However, for our purposes it will be
important to note that the Killing spinor defines a privileged
D=11 $SU(5)$ structure which satisfies certain differential
conditions and these are non-trivial even in the topologically
trivial neighbourhood. It should also be noted that the Killing
spinor defines a global $SU(5)$ structure if we restrict to
regions in which the Killing vector is time-like, which could in
fact be the whole manifold, if there are no points in which the
Killing vector becomes null.

The time-like case is analysed in detail in section 4. Working in
a neighbourhood where $K$ is time-like, we use the Killing spinor
equation and the $SU(5)$ structure to determine the general local
form of the geometry. Since $K$ is Killing the $SU(5)$ structure
in eleven-dimensions turns out to be mostly specified by an
$SU(5)$ structure in the ten-dimension base space orthogonal to
the orbits of the Killing vector. The only restriction on the
ten-dimensional $SU(5)$ structure is that the class
$\mathcal{W}_5$, defined later as the Lee-form of the (5,0)-form,
is exact and related to the norm of the Killing vector. We find
that, much, but not all, of the form of the geometry is determined
by the $SU(5)$ structure. In particular, there is a component of
the four-form field strength which is undetermined. The reason for
this fact is simply that this component drops out of the Killing
spinor equation. The necessary and sufficient form of the geometry
is presented in \p{sumone}, \p{sumtwo} and \p{sumthree}. By
analysing integrability conditions for the Killing spinor
equation, we also determine the extra constraints imposed on those
geometries admitting Killing spinors in order that they solve the
equations of motion. The extra constraints are presented in
\p{bone} and \p{btwo}. The equations of motion constrain the
component of the four-form that is not determined by supersymmetry
alone. Our results allow us to obtain some vanishing theorems for
compactifications with flux (for other such theorems in D=10
supergravity with NS three-form flux only, and assuming a
restricted class of configurations, see
\cite{ivpapone,ivpaptwo,Gauntlett:2002sc}).

The results for the time-like case presented here, and the
analogous results for the null-case, which we leave to future
work, would classify the local form of the most general bosonic
supersymmetric solutions of D=11 supergravity. A refinement of
this classification would be to find the additional restrictions
placed on the geometries in order that they preserve more than one
supersymmetry, and we discuss how in principle this could be done.

In section 5 we use the results for the timelike case to obtain
some some new solutions. In \cite{Cvetic:2000mh,Cvetic:2000db}
(see also \cite{Becker:1996gj,Duff:1997hf,Hawking:1997bg}) it was
shown that the membrane solution with a transverse manifold of
$SU(4)$ holonomy can be resolved by switching on additional
four-form flux via a harmonic four-form. Here we will show that
one can extend these solutions to include rotation. In
\cite{Gauntlett:2002nw} a D=5 generalisation of the G\"odel
solution was constructed. It was shown that it can be uplifted to
D=11 where it then preserves 5/8 supersymmetry. The topology of
the space is $\bR^{11}$ and there is a rotational one-form that
lives in an $\bR^4$ factor. We will show that there are further
solutions with more complicated rotation one-forms.

Section 6 briefly concludes.

\section{Killing spinors and differential forms}

The bosonic fields of D=11 supergravity consist of a metric, $g$,
and a three-form potential $A$ with four-form field strength
$F=dA$. The action for the bosonic fields is given by \bea
S=\frac{1}{2\kappa^2}\int d^{11} x {\sqrt{-g}}R
-\frac{1}{2}F\wedge *F - \frac{1}{6}C\wedge F\wedge F \eea where
$F=dC$. The equations of motion are thus given by \bea
R_{\mu\nu}-\frac{1}{12}(F_{\mu
\s_1\s_2\s_3}F{_{\nu}}{^{\s_1\s_2\s_3}}-
\frac{1}{12}g_{\mu\nu}F^2)&=&0\nn d*F+\frac{1}{2}F\wedge F&=&0
\eea We are interested in bosonic solutions to the equations of
motion that preserve at least one supersymmetry i.e. solutions
that admit at least one Killing spinor, $\epsilon$, which solves
\begin{equation}\label{killing}
\nabla_\mu\e+\frac{1}{288}[\Gamma{_\mu}{^{\nu_1\nu_2\nu_3\nu_4}}
-8\delta{_\mu^{\nu_1}}\Gamma^{\nu_2\nu_3\nu_4}]F_{\nu_1\nu_2\nu_3\nu_4}\e=0
\end{equation}
Note that due to the presence of the
terms involving the four-form, the supercovariant
derivative appearing in \p{killing} takes values in the Clifford
algebra and not just the spin subalgebra. Our conventions are
outlined in appendix A.

Note that in M-theory, the field-equation for the four-form
receives higher order gravitational corrections
\cite{vafawitten,duffetal}:
\be
d*F+\frac{1}{2}F\wedge F=-\beta X_8~,
\ee
where
\be
X_8=\frac{1}{192}(p_1^2-4p_2)~,
\ee
$\beta = 2\pi/T_5$ where $T_5=(2\pi)^{1/3}/(2\kappa^2)^{2/3}$ 
is the tension of a single fivebrane, and
the Pontryagin-forms are given by
\bea
p_1&=&-\frac{1}{8\pi^2}trR^2\nn
p_2&=&-\frac{1}{64\pi^4}tr R^4+\frac{1}{128\pi^4}(tr R^2)^2~.
\eea
Since most of our analysis
concerns the Killing spinor equation \p{killing} this correction
will not play a large role in the following.

Consider a geometry that admits  $N$ Killing spinors $\e^i$,
$i=1,..,N$. We can define the following  one, two and five-forms
that are symmetric in ${i,j}$: \bea
\K^{ij}_\mu&=&\bar\e^i\Gamma_\mu\e^j  \nn
\Omega^{ij}_{\mu_1\mu_2}&=&\bar\e^i\Gamma_{\mu_1\mu_2}\e^j\nn
\Sigma^{ij}_{\mu_1\mu_2\mu_3\mu_4\mu_5}&=&\bar\e^i\Gamma_{\mu_1\mu_2\mu_3\mu_4\mu_5}\e^j
\eea We can also define zero, three and four-forms which are
anti-symmetric in ${i,j}$ \bea X^{ij}&=&\bar\e^i\e^j \nn
Y^{ij}_{\mu_1\mu_2\mu_3}&=&\bar\e^i\Gamma_{\mu_1\mu_2\mu_3}\e^j\nn
Z^{ij}_{\mu_1\mu_2\mu_3\mu_4}&=&\bar\e^i\Gamma_{\mu_1\mu_2\mu_3\mu_4}\e^j
\eea Note that that the diagonal forms with $i=j$ are
non-vanishing if and only if $\e$ is. Indeed, in our conventions
the charge conjugation matrix is given by $C=\Gamma_0$, so the
zeroth component of the vector $K$ in an orthonormal frame is
given by $K^{ii}_0=-(\e^i)^T (\e^i)$. {}From the results presented
in the next sub-section, it similarly follows that $\Omega^{ii},
\Sigma^{ii}$ are also non-vanishing if and only if $\e$ is.

In fact, following \cite{harvey}, we can argue that there are
actually no points in the spacetime where $K\equiv K^{ii}$ and
hence $\e^i$ vanishes. We will use two facts which will be discussed
later: firstly that $K$ is either timelike or null and secondly
that $K$ is Killing. Suppose then, that $K$ and $\e^i$ vanish at a
point $p$. Consider a time-like geodesic through $p$ with tangent
vector $U$. Since $K$ is Killing, $U\cdot K$ is constant along the
geodesic and since it vanishes at $p$ it must in fact vanish along
the whole of the geodesic. However, $U$ is timelike while $K$ is
non-spacelike and hence $K$ and thus $\e^i$ must vanish along the
whole of the geodesic. As this argument applies to any timelike
geodesic passing through $p$ we conclude that $\e^i$ vanishes to
the future and past of the point $p$. Assuming analyticity, we
conclude that $\e^i$ vanishes everywhere (assuming that the
spacetime is connected) which contradicts the assumption that we
have a Killing spinor.

\subsection{Algebraic Relations}
The differential forms defined above are not all independent. They
satisfy certain algebraic relations which are a consequence of the
underlying Clifford algebra. One way of obtaining these is by
repeated use of {}Fierz identities. Another approach will be
mentioned later. Let us  illustrate this by considering the case
$i=j$ and dropping the $ij$ indices, which covers the most general
case when there is only one Killing spinor.

We first relate $\Omega^2$ and $\Sigma^2$ to $\K^2$. We use here
the convention that for any p-form $\alpha$ we have:
\[
\alpha^2\equiv\frac{1}{p!}
\alpha_{\mu_1\mu_2...\mu_p}\alpha^{\mu_1\mu_2...\mu_p}
\]
By performing {}Fierz rearrangements on $\K^2$, $\Omega^2$ and
$\Sigma^2$ in turn we find three linearly dependent equations.
Solving them we find:
\begin{equation}\label{omsquare}
  \begin{array}{c}
     \Sigma^2=-6\K^2 \\
    \Omega^2=-5\K^2 \
  \end{array}
\end{equation}
We also find the following relations:
\begin{equation}\label{complex}
\Omega{_{\mu_1}}{^{\sigma_1}}\Omega{_{\sigma_1}}{^{\nu_1}}
=-\K_{\mu_1}\K^{\nu_1}+\delta{_{\mu_1}}{^{\nu_1}}\K^2
\end{equation}
\begin{eqnarray}\label{sigmasigma}
\frac{1}{4!}\Sigma{_{\mu_1}}{^{\sigma_1\sigma_2\sigma_3\sigma_4}}
\Sigma{_{\sigma_1\sigma_2\sigma_3\sigma_4}}{^{\nu_1}}
&=&14\K_{\mu_1}\K^{\nu_1}-4\delta{_{\mu_1}}{^{\nu_1}}K^2
\end{eqnarray}
\begin{equation}\label{ikom}
i_{\K}\Omega=0
\end{equation}
\begin{equation}\label{iks}
i_{\K}\Sigma=\frac{1}{2}\Omega\wedge\Omega
\end{equation}
\begin{equation}\label{ikstars}
\K^{\s}(*\Sigma)_{\s\nu_1\nu_2\nu_3\nu_4\nu_5}
=\Omega{_{\nu_1}}{^{\s}}\Sigma_{\s\nu_2\nu_3\nu_4\nu_5}
\end{equation}
\bea K^2\Omega\wedge\Sigma=\frac{1}{2}
\K\wedge\Omega\wedge\Omega\wedge\Omega \eea These are by no means
exhaustive.

\subsection{Differential Relations}
The covariant derivatives of the differential forms can be
calculated by using the fact that a Killing spinor satisfies: \be
\overline{\nabla_\mu\e^i}=\frac{1}{288}\bar\e^i[\Gamma{_\mu}{^{\nu_1\nu_2\nu_3\nu_4}}
+8\delta{_\mu^{\nu_1}}\Gamma^{\nu_2\nu_3\nu_4}]F_{\nu_1\nu_2\nu_3\nu_4}
\ee We find \bea\label{cov} \nabla_\mu
K^{ij}_{\nu}&=&\frac{1}{6}\Omega^{ij}{^{\s_1\s_2}}F_{\s_1\s_2\mu\nu}
+\frac{1}{6!}\Sigma^{ij}{^{\s_1\s_2\s_3\s_4\s_5}}*F_{\s_1\s_2\s_3\s_4\s_5\mu\nu}\nn
\nabla_{\mu}\Omega^{ij}_{\nu_1\nu_2}&=&\frac{1}{3\cdot
4!}g_{\mu[\nu_1}\Sigma^{ij}{_{\nu_2]}}{^{\s_1\s_2\s_3\s_4}}F_{\s_1\s_2\s_3\s_4}+\frac{1}{3\cdot
3!}\Sigma^{ij}{_{\nu_1\nu_2}}{^{\s_1\s_2\s_3}}F_{\mu\s_1\s_2\s_3}\nn
&-&\frac{1}{3\cdot
3!}\Sigma^{ij}{_{\mu[\nu_1}}{^{\s_1\s_2\s_3}}F_{\nu_2]\s_1\s_2\s_3}+\frac{1}{3}\K^{ij~\s}F_{\s\mu\nu_1\nu_2}\nn
\nabla_{\mu}\Sigma^{ij}_{\nu_1\nu_2\nu_3\nu_4\nu_5}&=&
\frac{1}{6}{\K^{ij~\s}}*F_{\sigma\mu\nu_1\nu_2\nu_3\nu_4\nu_5}
-\frac{10}{3}F_{\mu[\nu_1\nu_2\nu_3}\Omega^{ij}_{\nu_4\nu_5]}-\frac{5}{6}F_{[\nu_1\nu_2\nu_3\nu_4}\Omega^{ij}_{\nu_5]\mu}\nn
&-&\frac{10}{3}g_{\mu[\nu_1}\Omega^{ij}{_{\nu_2}}{^{\sigma}}F_{|\sigma
|\nu_3\nu_4\nu_5]}
+\frac{5}{6}F_{\mu[\nu_1|\sigma_1\sigma_2|}(*\Sigma^{ij}){^{\sigma_1\sigma_2}}{_{\nu_2\nu_3\nu_4\nu_5]}}\nn
&+&\frac{5}{6}F_{[\nu_1\nu_2|\sigma_1\sigma_2|}(*\Sigma^{ij}){^{\sigma_1\sigma_2}}{_{\nu_3\nu_4\nu_5]\mu}}
-\frac{5}{9}g_{\mu[\nu_1}F_{\nu_2
|\sigma_1\sigma_2\sigma_3|}(*\Sigma^{ij}){^{\sigma_1\sigma_2\sigma_3}}{_{\nu_3\nu_4\nu_5]}}\nn
\eea The exterior derivatives of the forms are thus given by
\bea\label{dk}
d\K^{ij}&=&\frac{2}{3}i_{\Omega^{ij}}F+\frac{1}{3}i_{\Sigma^{ij}}*F\\
\label{dktwo}
d\Omega^{ij}&=&i_{\K^{ij}}F\\
\label{dkthree} d\Sigma^{ij}&=&i_{\K^{ij}}* F -\Omega^{ij}\wedge F
\eea where eg $(i_\Omega F)_{\mu\nu}=(1/2!)\Omega^{\rho_1\rho_2}
{}F_{\rho_1\rho_2\mu\nu}$.

{}From the first equation in \p{cov} we can immediately deduce the
important result that each of the $\K^{ij}$ are Killing vectors.
Moreover, using the Bianchi identity, it is simple to show that
\be\label{ksym} \CL_{K^{ij}} F=0 \ee for any $\K^{ij}$. Thus any
geometry $(g,F)$ admitting Killing spinors posesses symmetries
generated by $\K^{ij}$.

Notice, as somewhat of an aside, that using \p{dk} we also have
\be\label{somewhat}
\CL_{K^{ij}}*F=i_{K^{ij}}(d*F+\frac{1}{2}F\wedge F) \ee Now the
fact that $K^{ij}$ is Killing and the condition \p{ksym} implies
that both the left and right hand side must vanish separately.
This means that the presence of a Killing spinor implies that some
components of the equation of motion for the four-form are
automatically satisfied\footnote{Note that we are ignoring the
$X_8$ correction here. To fully consistently incorporate it one
needs to also consider higher order correction terms to the
Killing spinor equation. However, we note here that if $i_{K^{ij}}
X_8=0$ the above equation is consistent.}. Notice also that this
calculation provides a check on the sign appearing in the
Chern-Simons term in the D=11 supergravity Lagrangian, given the
form of the Killing spinor equation and the conventions for the
Clifford algebra.

We next note that \p{dktwo} is strikingly similar to the notion of
generalised calibration for static membranes introduced in
\cite{gpt} following \cite{Gutowski:1999iu}. Indeed consider the
special case that $i=j$ and when $K=K^{ii}$ is a static Killing
vector. Then taking into account \p{ikom} we see that \p{dktwo} is
exactly the same equation satisfied by a generalised calibration
$\Omega$ for a membrane that was introduced in \cite{gpt}. Recall
that the significance of generalised calibrations is that they
calibrate supersymmetric brane world-volumes in the presence of
non-zero four-form flux. What we have shown here is that
supersymmetric D=11 geometries automatically give rise to
generalised calibrations $\Omega$.

That one gets the same result, in this special case, either from
D=11 supergravity or from the world-volume theory using
kappa-symmetry as in  \cite{gpt}, is perhaps not that surprising
since it is well known that the kappa-symmetry of the
super-membrane implies the equations of motion of D=11
supergravity \cite{bst1,bst2}. What is particularly inteteresting,
though, is that the D=11 supergravity result indicates that the
notion of generalised calibrations might be extended to more
general settings than that studied in \cite{gpt}. {}Firstly, since
\p{dktwo} is valid when $K$ is not only static but more generally
stationary, it suggests that the analysis of \cite{gpt} can be
straightforwardly extended to the stationary case, as assumed in
that paper. Secondly, \p{dktwo} is also valid when when $K$ is
null and it should be very interesting to elucidate the
physical interpretation of this from the world-volme point of
view. {}Finally, when there is more than one Killing spinor,
$K^{ij}$ with $i\ne j$ can also be spacelike.
This latter case is at least partially related to the issue
discussed at the end of section II of \cite{gpt} concerning the
fact that static supersymmetric branes can have some flat
directions.

The notion of generalised calibrations for fivebranes is more
complicated due to the fact that the fivebrane world-volume has a
self-dual three-form, which is responsible for the fact that
membranes can end on fivebranes. An initial investigation was
undertaken in \cite{blw} for the case of static configurations,
where it was argued that the generalised calibration for the
five-brane is a pair consisting of a spatial five-form and
two-form. {}For static $K^{ii}$ these correspond to the spatial
part of $\Sigma$ and $\Omega$. The possibility of the five-form
not being closed was considered in \cite{blw} and argued to be
related to Wess-Zumino terms in the fivebrane worldvolume theory.
That $\Omega$ might also not be closed was not considered in
\cite{blw}, but here we see that this is the general situation and
it is not difficult to see that this is again related to
Wess-Zumino terms in the fivebrane worldvolume theory. Moreover,
our analysis reveals the correct differential expressions when
$K^{ii}$ is stationary and also when it is null, the latter case
again being particularly intriguing. As for membranes,
there also seems to be an interesting generalisation
for spacelike $K^{ij}$ with $i\ne j$.

Returning to \p{cov}, it is also useful to note that we can
extract \bea\label{eqn:dstarOmega}
(*d*\Omega^{ij})_\nu&=&-\frac{1}{3\cdot 4!}(\Sigma^{ij})_\nu{}
^{\s_1\s_2\s_3\s_4}F_{\s_1\s_2\s_3\s_4}\nn
(*d*\Sigma^{ij})_{\nu_1\nu_2\nu_3\nu_4} &=&
\frac{8}{3}(\Omega^{ij})^\s{}_{[\nu_1}F_{\nu_2\nu_3\nu_4]\s}
+\frac{2}{9} (*\Sigma^{ij})^{\s_1\s_2\s_3}{}_{[\nu_1\nu_2\nu_3}
{}F_{\nu_4]\s_1\s_2\s_3} \eea {}Finally, using the algebraic
results of the last subsection it is simple to conclude that the
Lie-derivative of $\Omega$ and $\Sigma$ with respect to $K$
vanish: \bea\label{liederstr} \CL_K \Omega&=&0\nn \CL_K \Sigma&=&0
\eea

The corresponding equations for $X,Y$ and $Z$ are given by \bea
\nabla_\mu X^{ij}&=&-\frac{1}{3\cdot
3!}(Y^{ij})^{\rho_1\rho_2\rho_3}F_{\rho_1\rho_2\rho_3\mu_1}\nn
\nabla_\mu
Y^{ij}_{\nu_1\nu_2\nu_3}&=&-\frac{1}{3}X^{ij}F_{\mu\nu_1\nu_2\nu_3}-\frac{1}{6\cdot
3!}Y^{ij}{^{\rho_1\rho_2\rho_3}}*F_{\rho_1\rho_2\rho_3\mu\nu_1\nu_2\nu_3}\nn
&-&\frac{1}{4}Z^{ij}{^{\rho_1\rho_2}}{_{\mu[\nu_1}}F_{\nu_2\nu_3]\rho_1\rho_2}
-\frac{1}{2}Z^{ij}{^{\rho_1\rho_2}}{_{[\nu_1\nu_2}}F_{\nu_3]\mu\rho_1\rho_2}
-\frac{1}{6}g_{\mu[\nu_1}Z^{ij}_{\nu_2}{^{\rho_1\rho_2\rho_3}}F_{\nu_3]\rho_1\rho_2\rho_3}\nn
\nabla_\mu
Z^{ij}_{\nu_1\nu_2\nu_3\nu_4}&=&\frac{2}{3}Y^{ij}{_{\mu[\nu_1}}{^{\rho}}F_{\nu_2\nu_3\nu_4]\rho}
-2Y^{ij}{_{[\nu_1\nu_2}}{^{\rho}}F_{\nu_3\nu_4]\mu\rho}\nn
&-&g_{\mu[\nu_1}Y^{ij}{_{\nu_2}}{^{\rho_1\rho_2}}F_{\nu_3\nu_4]\rho_1\rho_2}
-\frac{1}{9}*Z^{ij}{_{\mu[\nu_1\nu_2\nu_3}}{^{\rho_1\rho_2\rho_3}}F_{\nu_4]\rho_1\rho_2\rho_3}\nn
&+&\frac{1}{18}*Z^{ij}{_{\nu_1\nu_2\nu_3\nu_4}}{^{\rho_1\rho_2\rho_3}}F_{\mu\rho_1\rho_2\rho_3}
+\frac{1}{36}g_{\mu[\nu_1}*Z^{ij}{_{\nu_2\nu_3\nu_4]}}{^{\rho_1\rho_2\rho_3\rho_4}}F_{\rho_1\rho_2\rho_3\rho_4}\nn
\eea and \bea\label{dx} (dX^{ij})_{\mu_1}&=&-\frac{1}{3\cdot
3!}(Y^{ij})^{\rho_1\rho_2\rho_3}F_{\rho_1\rho_2\rho_3\mu_1}\nn
(dY^{ij})_{\mu_1\mu_2\mu_3\mu_4}&=&-\frac{1}{9}(Y^{ij})^{\rho_1\rho_2\rho_3}
(*F)_{\rho_1\rho_2\rho_3\mu_1\mu_2\mu_3\mu_4}+(Z^{ij}){_{[\mu_1\mu_2}}
{^{\rho_1\rho_2}}F_{\mu_3\mu_4]\rho_1\rho_2}\nn
&-&\frac{4}{3}X^{ij}F_{\mu_1\mu_2\mu_3\mu_4}\nn
(dZ^{ij})_{\mu_1\mu_2\mu_3\mu_4\mu_5}&=&
-\frac{20}{3}(Y^{ij}){_{[\mu_1\mu_2}}{^{\rho_1}}F_{\mu_3\mu_4\mu_5]\rho_1}
-\frac{5}{18}(*
Z^{ij}){_{[\mu_1\mu_2\mu_3\mu_4}}{^{\rho_1\rho_2\rho_3}}
{}F_{\mu_5]\rho_1\rho_2\rho_3}\nn \eea and also \bea
(*d*Y^{ij})_{\nu_1\nu_2}&=&0\nn
(*d*Z^{ij})_{\nu_1\nu_2\nu_3}&=&-\frac{1}{2}
(Y^{ij})^{\s_1\s_2}{}_{[\nu_1}F_{\nu_2\nu_3]\s_1\s_2}
-\frac{1}{36}(*Z^{ij})_{\nu_1\nu_2\nu_3}{}^{\s_1\s_2\s_3\s_4}
{}F_{\s_1\s_2\s_3\s_4}\nn \eea

\subsection{Integrability}

The integrability of the Killing spinor equation allows us to
relate geometries admitting Killing spinors to those that in
addition solve the equations of motion. As shown in the appendix,
integrability of the Killing spinor equation implies that
\bea\label{rrghy} 0&=&[R_{\mu\nu}-\frac{1}{12}(F_{\mu
\s_1\s_2\s_3}
{}F{_{\nu}}{^{\s_1\s_2\s_3}}-\frac{1}{12}g_{\mu\nu}F^2)]\Gamma^{\nu}\e^i\nn
&-&\frac{1}{6\cdot 3!}*(d*F+\frac{1}{2}F \wedge
{}F)_{\s_1\s_2\s_3}(\Gamma_{\mu}{^{\s_1\s_2\s_3}}-6\delta^{\s_1}_{\mu}\Gamma^{\s_2\s_3})\e^i\nn
&-&\frac{1}{6!}dF_{\s_1\s_2\s_3\s_4\s_5}(\Gamma_{\mu}{^{\s_1\s_2\s_3\s_4\s_5}}
-10\delta_{\mu}^{\s_1}\Gamma^{\s_2\s_3\s_4\s_5})\e^i \eea for each
Killing spinor $\epsilon^i$.

Assume that we have a geometry $(g,F)$ that admits Killing spinors
and that also solves the Bianchi identity and the equations of
motion for the four-form $F$. We then deduce from \p{rrghy} that
\be\label{int} 0=E_{\mu\nu} \Gamma^\nu\epsilon^i=0 \ee where
$E_{\mu\nu}=0$ is equivalent to the Einstein equations. We now
follow the analysis of \cite{Gauntlett:2002nw}: hitting \p{int}
with $\bar\e^i$ we conclude that \be\label{intone} E_{\mu\nu}
K^{\nu}=0 \ee On the other hand if we hit it with
$E_{\mu\sigma}\Gamma^\sigma$ we conclude that \be\label{inttwo}
E_{\mu\nu}E_\mu{}^\nu=0\qquad {\rm no\quad sum\quad on}\quad \mu
\ee

As we shall discuss, the Killing vector $K\equiv K^{ii}$ is either
a timelike or null. We first assume that it is timelike.
Introducing an orthonormal frame with $K=e^0$, we deduce from
\p{intone} that $E_{\mu0}=0$. The indices in \p{inttwo} then run
over spatial indices only and we conclude that $E_{\mu\nu}=0$
since there are no non-trivial null vectors in a euclidean space.
Alternatively if $K$ is null we can set up a D=11 frame \be ds^2=
2e^+ e^- +e^a e^a \ee for $a=1,\dots 9$, with $K=e^+$. Now
\p{intone} implies $E_{-\mu}=0$ while \p{inttwo} implies
$E_{+a}=E_{ab}=0$. Hence, one just needs to impose $E_{++}=0$ to
obtain a full supersymmetric solution.

These results have some obvious practical benefits in finding
explicit solutions.

\section{Classifying solutions using $G$-structures}

In the last section we derived a number of necessary conditions,
both algebraic and differential, for a geometry to possess Killing
spinors. A useful organisational principle is that of a
$G$-structure.

Let us begin by recalling the definition of $G$-structure of a
$n$-dimensional manifold $M$. The frame bundle is a principal
$Gl(n)$ bundle and a $G$-structure is simply a principal
$G$-sub-bundle. Often, the $G$-structure can be equivalently
specified by the existence of no-where vanishing $G$-invariant
tensors, and it is in this geometric guise that $G$-structures
will be important here. {}For example, a metric of euclidean
signature is equivalent to an $O(n)$ structure and if supplemented
with an orientation is equivalent to an $SO(n)$ structure. An
almost complex structure $J$ is equivalent to a $Gl(n/2,C)$
structure, and if supplemented with an hermitian metric is
equivalent to a $U(n/2)$ structure, and so on. In D=11
supergravity the manifolds are equipped with a Lorentzian metric,
and a spin structure, so the frame bundle can always be reduced to
$Spin(10,1)$ and hence there is always a $Spin(10,1)$ structure.

Let us explain the main ideas in classifying $G$-structures using
$G\subset Spin(10,1)$ as an example (see e.g.
\cite{sal1,sal2,sal3} for further discussion). Consider a
$G\subset Spin(10,1)$ structure specified by $G$-invariant tensors
and/or spinors, that we collectively define by $\eta$. The
essential idea is simple: as $G$ defines a metric, one can take
the covariant derivative of $\eta$ with respect to the Levi-Civita
connection and then decompose the result into irreducible
$G$-modules. In more detail, one first uses the fact that there is
no obstruction to finding a connection preserving the structure.
If we choose one, $\nabla'$, then one notes that $\nabla
\eta=(\nabla -\nabla')\eta$. Now $(\nabla -\nabla')$ is a tensor
with values in $T^*\otimes spin(10,1)$ but acting on the
$G$-invariant $\eta$ we see that the piece of  $\nabla \eta$ that
is independent of $\nabla'$ is  given by an element of $T^*\otimes
g^\perp$ where $g\oplus g^\perp =spin(10,1)$. This element is
known as the intrinsic torsion and can be decomposed into
irreducible $G$-modules: $\nabla\eta \leftrightarrow T^*\otimes
g^\perp=\oplus_i \mathcal W_i$. In one extreme, all of these
modules $\mathcal W_i$ are present and one has the most general
type of $G$-structure. In the other extreme, all of the modules
vanish and the tensors are covariantly constant giving rise to
manifolds with special holonomy $G$.

We can now use this language to interpret the algebraic and
differential conditions that we obtained for Killing spinor
bi-linears in the last section. In particular, it will provide us
with a framework for classifying the local form of all
supersymmetric solutions of D=11 supergravity.

Start with the most general supersymmetric geometry preserving (at
least) one Killing spinor. That is, start with a D=11 geometry
with a $Spin(10,1)$ structure and assume that we have a globally
defined spinor, which is equivalently specified by the tensors
$K,\Omega,\Sigma$ constructed from the bi-linears in the Killing
spinors that we introduced in the last section. We showed in
section 2 that the Killing spinor is non-vanishing. At a point,
the isotropy group of the spinor is known to be either $SU(5)$ or
$(Spin(7)\ltimes\bR^8)\times \bR$ \cite{bryant} depending on
whether $K$ is time-like or null, respectively. A spacelike $K$ is
not possible. The forms $K,\Omega,\Sigma$ are invariant under
$SU(5)$ or $(Spin(7)\ltimes\bR^8)\times \bR$ in each case, and one
can interpret the complete algebraic identities that
$K,\Omega,\Sigma$ satisfy, of which we obtained a subset in the
last section, as simply encoding this information.

In the ``null''-case $K$ is null everywhere and in the
``time-like'' case it is time-like at least at a point and hence
at least in a neighbourhood of such a point. In the null case the
Killing spinor, or equivalently $K,\Omega,\Sigma$, define a $D=11$
$(Spin(7)\ltimes\bR^8)\times \bR$ structure. A description of this
unusual structure can be found in \cite{bryant} (see also
\cite{Figueroa-O'Farrill:1999tx}).

For the time-like case, in situations in which $K$ is time-like
everywhere, the Killing spinor, or equivalently $K,\Omega,\Sigma$,
define a $D=11$ $SU(5)$ structure. It can happen, however, that
there are supersymmetric geometries in which $K$ is time-like in
some regions but becomes null at some points in the manifold, such
as at a horizon. In the generic case, therefore, we observe that
if we restrict to regions in which $K$ is time-like then
$K,\Omega,\Sigma$ still defines an $SU(5)$ structure. It is in
this sense that we will say that the timelike case has an $SU(5)$
structure. This will be sufficient for our purposes, but we
comment that since $(K,\Omega,\Sigma)$ are globally defined there
may be a another mathematical language to describe this situation.

The analysis in the next section for the time-like case, will be
carried out in a neighbourhood in which the Killing vector is
timelike. Since the neighbourhood is topologically trivial, the
frame bundle can always be trivialised. To avoid possible
confusion, we emphasise that the Killing spinor defines a
privileged D=11 $SU(5)$ structure which satisfies certain
differential conditions and these are non-trivial even in the
topologically trivial neighbourhood. For example the privileged
$SU(5)$ structure is equivalent to a certain metric on this
neighbourhood.

So any geometry admitting a D=11 Killing spinor will either have a
$SU(5)$ structure or a $(Spin(7)\ltimes\bR^8)\times \bR$ structure
that is built from the Killing spinor.
The next steps in classifying the local form of the
most general supersymmetric bosonic geometries are to
determine, in each case, the intrinsic
torsion of the $G$-structure i.e., the type of $G$-structure that
arises according to the classification of $G$-structures, and then
to see how much of the form of the geometry is specified by the
structure. This can be achieved by analysing the differential
conditions imposed on the tensors $K,\Omega,\Sigma$ that we
obtained from the Killing spinor equation earlier.

We will carry out this analysis in detail for the case of a single
timelike Killing spinor in the next section. We will see that the
differential conditions restrict the type of $SU(5)$ structure.
In addition we will determine the general local
form of the geometry and show that it is almost completely
determined by the $SU(5)$ structure: we will see that there is a
component of the four-form that is not fixed by supersymmetry
alone. We will also prove a converse result, namely that given
such a $SU(5)$ structure, with the appropriately specified
four-form, then the geometry does indeed admit at least one
Killing spinor.

As noted in the last section, for the timelike case the
integrability conditions for the Killing spinor imply that in
order to have a supersymmetric solution to the equations of
motion, one just needs to impose the Bianchi identity and the
equations of motion for the four-form. These conditions impose
further independent constraints and in particular constrain the
component of the four-form not specified by supersymmetry.

A similar analysis for the null case, which we will leave for
future work, would then complete a classification of the local
form of the most general types of bosonic D=11 supergravity
solutions. A finer classification would be to find the additional
restrictions placed on these geometries in order that they
preserve more than one supersymmetry. The additional Killing
spinors will either be time-like or null and hence there will be
several $SU(5)$ and/or $(Spin(7)\ltimes\bR^8)\times \bR$
structures. Equivalently, the structure group can be reduced
further to the maximal common subgroup which is the isotropy group
of all of the spinors. So a first step to obtain this refined
classification would be to classify all of the different isotropy
groups of 1,...,32 spinors. In principle, one way of tackling this
problem would be to derive algebraic conditions on the tensors
$K^{ij},\Omega^{ij},\Sigma^{ij}$ and $X^{ij},Y^{ij},Z^{ij}$ using
various {}Fierz identities. However, the calculations we carried
out for the tensors with $i=j$ in the last section were already
very involved and this would be a very clumsy approach. It should
be more efficient to generalise the work of \cite{bryant}. In
addition, we note that the groups appearing in tables 1 and 2 of
\cite{Figueroa-O'Farrill:1999tx} will certainly be relevant.

The second step in obtaining the refined classification would be
to determine the intrinsic torsion of the $G$-structure defined by
the Killing spinors and to see how much of the form of the
geometry is specified by the structure. Again this information is
contained in the differential conditions imposed on the tensors
$K^{ij},\Omega^{ij},\Sigma^{ij}$ and $X^{ij},Y^{ij},Z^{ij}$ that
we obtained from the Killing spinor equation. The results of the
next section on the case of a single time-like spinor are
encouraging that this entire programme could be carried out.

\section{The stationary case and $SU(5)$ structure}

In this section we will determine the most general form of
geometries admitting at least one timelike Killing spinor. The
spinor can be used to construct a one-form $K$, a two-form
$\Omega$ and a five-form $\Sigma$. Working in a neighbourhood
where $K$ is time-like, these forms together specify a privileged
$SU(5)$ structure in D=11. An important restriction on this D=11
structure is that the dual time-like vector field to $K$ is
Killing. We can thus introduce a time coordinate along the orbits
of the Killing vector, so that we have $K=-\Delta^2 (dt+\omega)$,
with $\Delta$ and $\omega$ independent of $t$. The metric then
takes the form: \bea\label{eqn:metric} ds_{11}^2=-\Delta^2
(dt+\omega)^2+\Delta^{-1}g_{mn}dx^m dx^n \eea and $K^2=-\Delta^2$.
The metric $\Delta^{-1}g_{mn}$ is a metric on the ten dimensional
euclidean spatial base manifold, which we will denote by $B$,
defined via the orthogonal projection of the eleven dimensional
metric with respect to the Killing vector. We will work in these
coordinates and then obtain the full geometry by analytic
continuation at the end.

{}From \p{ikom} and \p{liederstr} we immediately deduce that
$\Omega$ is a two-form on the base manifold. If we raise an index
using the metric $g$ we obtain an almost complex structure on $B$.
The metric $g$ is then hermitian with respect to this almost
complex structure and $\Omega$ is the K\"ahler form.

Using \p{iks} it follows that the five-form $\Sigma$ can be
written as \bea\label{sigmaequals}
\Sigma=\frac{1}{2}\Delta^{-1}e^0\wedge\Omega\wedge\Omega+\Delta^{-3/2}\chi
\eea where, again using \p{liederstr}, $\chi$ is a five-form on
$B$ and we have defined \be e^0=\Delta(dt +\omega) \ee which can
be used to build an orthonormal frame in $D=11$. If we now define
\bea \theta=\chi-i *\chi \eea where, in this section, $*$ is the
Hodge star with respect to the metric $g$, we conclude from
\p{ikstars} that $\theta$ is a $(5,0)$ form on $B$. This means
that the ten-dimensional base manifold $B$ admits an $SU(5)$
structure specified by $\Omega,\theta$, or equivalently by
$g,\Omega,\chi$. In most of the subsequent analysis, the focus
will be on the D=10 $SU(5)$ structure on $B$.

Note that the factors of $\Delta$ were inserted in the definition
of $\chi$ in \p{sigmaequals} to ensure that the $SU(5)$ structure
satisfies the compatibility condition
\begin{eqnarray}\label{ch6:compatibility}
\chi\wedge *\chi=-2^4\frac{ \Omega^5}{5!}
\end{eqnarray}
Note also that \bea \chi^2={16} \eea where the indices here are
contracted using the metric $g$, which can be deduced from
\p{omsquare}.

The existence of an $SU(5)$ structure allows us to decompose the
complexified space of forms on $B$ into irreducible
representations of $SU(5)$, and this will be very useful in the
following. We first decompose the space of forms into
$(p,q)$-forms. Pure forms of type $(p,0)$ form irreducible
representations of $SU(5)$. {}For mixed forms we need to remove
traces taken with $\Omega$ to form irreducible representations so
these split further into: \bea\label{eqn:reps}
\Lambda^{(1,1)}&\cong& \Lambda^{(1,1)}_{0}\oplus{\mathbb{R}}\nn
\Lambda^{(2,1)}&\cong& \Lambda^{(2,1)}_0\oplus \Lambda^{(1,0)}\nn
\Lambda^{(2,2)}&\cong& \Lambda^{(2,2)}_0\oplus
\Lambda^{(1,1)}_0\oplus{\mathbb{R}} \nn \Lambda^{(3,1)}&\cong&
\Lambda^{(3,1)}_0\oplus \Lambda^{(2,0)}\nn \Lambda^{(3,2)}&\cong&
\Lambda^{(3,2)}_0\oplus \Lambda^{(2,1)}_0\oplus \Lambda^{(1,0)}\nn
\Lambda^{(4,1)}&\cong&\Lambda^{(4,1)}_0\oplus \Lambda^{(3,0)} \eea
where the subscript $0$ denotes a traceless form. The rest are
determined by complex conjugation and by noting that * maps a
$(p,q)$-form to $(5-q,5-p)$-form.

It will be helpful at this point to briefly review the
classification of $SU(5)$ structures on ten-dimensional Riemannian
manifolds (further comments are made in appendix C). We noted in
the last section that $G$-structures are classified by the
intrinsic torsion, which is an element of $T^*\otimes g^\perp$.
Here $g^\perp$ is defined by $su(5)\oplus g^\perp\cong so(10)$.
Noting that the adjoint of $so(10)$ decomposes under $su(5)$ via
${\bf 45}\to {\bf 1}+{\bf 10}+{\bf \bar {10}}+{\bf 24}$ and that
${\bf 24}$ is the adjoint of $su(5)$, we conclude that the
intrinsic torsion is given by the $SU(5)$ modules:
\be\label{ordermod} ({\bf 5}+{\bf \bar{5}})\times ({\bf 1}+{\bf
10}+{\bf \bar{10}})\to ({\bf 10}+{\bf \bar{10}})+ ({\bf 40}+{\bf
\bar{40}})+ ({\bf 45}+{\bf \bar{45}})+ ({\bf
5}+{\bf\bar{5}})+({\bf 5'}+{\bf \bar{5}'}) \ee In other words, the
intrinsic torsion is given by five $SU(5)$ modules: $T^*\otimes
g^\perp\simeq \mathcal{W}_1\oplus\mathcal{W}_2\oplus\mathcal{W}_3
\oplus\mathcal{W}_4\oplus\mathcal{W}_5$, where
conventionally\footnote{The modules
$\mathcal{W}_1\oplus\mathcal{W}_2\oplus\mathcal{W}_3\oplus\mathcal{W}_4$
are those arising in the classification of $U(5)$ structures
\cite{grayh}.} the $\mathcal{W}_i$ are given in the order noted in
\p{ordermod}. The component of the intrinsic torsion in the module
$\mathcal{W}_i$ will be denoted by $W_i$.

It will be very important in the following to use the fact that
the $W_i$, and hence the intrinsic torsion, are determined by
$d\Omega$ and $d\chi$. One sees that this is possible by
consideration of the $su(5)$ irreps appearing in $d\Omega$ and
$d\chi$.  Consider first the three-form $d\Omega$ corresponding to
the ${\bf 120}$ of $SO(10)$. Since $\Omega$ is a $(1,1)$ form,
$d\Omega$ will have a $(3,0)+(0,3)$ piece and also a $(2,1)+(1,2)$
piece. Removing the trace from the latter pieces, one obtains the
decomposition ${\bf 120}\to{\bf 45}+{\bf \bar{45}}+{\bf 10}+{\bf
\bar{10}}+{\bf 5}+{\bf \bar{5}}$ under $SU(5)\subset SO(10)$.
Similarly, since $\chi$ is the real part of a $(5,0)$ form, the
six-form $d\chi$ will have a $(5,1)+(1,5)$ and a $(4,2)+(2,4)$
part. These give rise to the representations
$\mathbf{5}+\bar{\mathbf{5}}+
\mathbf{10}+\bar{\mathbf{10}}+\mathbf{40}+\bar{\mathbf{40}}$. We
thus see that $d\Omega$ and $d\chi$ contain all the irreps
appearing in the $W_i$. In more detail we can define the following
irreducible components of $d\Omega$ and $d\chi$:
\begin{eqnarray}\label{thedubi}
&\chi\wedge d\Omega =\Omega\wedge d\chi=W_1\wedge
\frac{\Omega^3}{3!}\nn &(d\chi)^{4,2}+(d\chi)^{2,4}=W_2\wedge
\Omega+\frac{1}{3}W_1\wedge \frac{\Omega^2}{2!}\nn
&(d\Omega)^{2,1}+(d\Omega)^{1,2}=W_3+\frac{1}{4}W_4\wedge\Omega
\nn &W_4=\Omega\lrcorner d\Omega \nn   &W_5=\chi\lrcorner d\chi
\end{eqnarray}
with $W_1=*(\Omega\wedge d\chi)$. Here we have introduced the
notation $\omega\lrcorner\nu$ which contracts a $p$-form $\omega$
into a $n+p$-form $\nu$ via: \be (\omega \lrcorner \nu)_{i_1\dots
i_n}= \frac{1}{p!}\omega^{j_1\dots j_p} \nu_{j_1\dots j_pi_1\dots
i_n} \ee A more precise connection between the intrinsic torsion
and the $W_i$ is presented in appendix C. Note that the
$\mathbf{10}+\bar{\mathbf{10}}$ part of $d\chi$ is related to the
$\mathbf{10}+\bar{\mathbf{10}}$ of $d\Omega$ through the condition
$\Omega\wedge\chi=0$. {}For orientation, note that the almost
complex structure is integrable iff $W_1=W_2=0$ so that manifolds
with an $SU(5)$-structure of type $\mathcal{W}_3
\oplus\mathcal{W}_4\oplus\mathcal{W}_5$ are hermitian manifolds.
Also if all $W_i$ vanish so that the intrinsic torsion vanishes
then the manifold is Ricci flat and has holonomy $G\subseteq
SU(5)$. We will see that the $SU(5)$ structure arising on the base
manifold $B$ is only weakly restricted in general.

We are now ready to relate the components of $F$ to the
$SU(5)$-structure $(g,\Omega,\chi)$. We will see that almost all
of $F$ is determined by the structure. {}First we write
\begin{equation}\label{ch6:F}F
=\Delta^{-1}e^0\wedge G+H
\end{equation}
where $G$ is a three-form and $H$ is a four-form defined on $B$.
The eleven dimensional Hodge dual of $F$ is thus given by:
$$*_{11}F=\Delta^{-3}*G+\Delta^{-1}e^0\wedge *H$$
We find from \p{dktwo} that \bea\label{eqn:G} G=d\Omega \eea and
from \p{dkthree} that \bea\label{eqn:detrH}
d(\Delta^{-3/2}\chi)+\frac{1}{2} d\omega\wedge\Omega\wedge\Omega=
*H-\Omega\wedge H \eea

So $G$ is clearly determined by the structure. We now attempt to
solve \p{eqn:detrH} for $H$. Introducing the map \bea
\Theta:\Lambda^4(B)\rightarrow\Lambda^4(B)\nn \a\mapsto
\a-*(\Omega\wedge \a) \eea we can rewrite \p{eqn:detrH} as:
\bea\label{thetaH}
*\Theta(H)=d(\Delta^{-3/2}\chi)+\frac{1}{2}d\omega\wedge\Omega\wedge\Omega
\eea Now $H$ is a four-form and can be split into irreducible
$SU(5)$ representations. The $\mathbf{210}$ of $SO(10)$ decomposes
under $SU(5)$ as,
$$\mathbf{210}\rightarrow
\mathbf{1}+\mathbf{5}+\bar{\mathbf{5}}+\mathbf{10}+\bar{\mathbf{10}}
+\mathbf{24}+\mathbf{40}+\bar{\mathbf{40}}+\mathbf{75}$$ We can
thus write \be H=H_1+H_{5+\bar 5}+H_{10+\bar {10}}
+H_{24}+H_{40+\bar {40}}+H_{75} \ee Each four-form $H_i$ can then
be written in terms of certain $(p,q)$ forms defining $SU(5)$
irreps. Using the identities listed in \p{sufids} a calculation
shows that each irreducible representation is an eigenvector of
$\Theta$ with eigenvalues given in table \ref{table:maptheta}.
Note that three of the representations have zero eigenvalue.
This has important consequences, as we shall see.
We next split the
right hand side of \p{thetaH} into irreducible $SU(5)$ components.
{}For $d\chi$ this was noted above. For the second term we write
$d\omega=d\omega^{(0)}\Omega+d\omega^{(1,1)}_0+d\omega^{(2,0)}+d\omega^{(0,2)}$,
corresponding to the decomposition ${\bf 45}\to {\bf 1}+{\bf
24}+{\bf 10}+{\bf \bar {10}}$.

\begin{table}[tb]
\begin{center}
\begin{tabular}{|c|c|}
  \hline
  Irreps  & E-Values \\ \hline
  ${\bf 1}$ & $-2$ \\ \hline
  ${\bf 5},{\bf \bar 5}$ & $0$ \\ \hline
  ${\bf 10}, {\bf \bar {10}}$ & $-1$ \\ \hline
  ${\bf 24}$ & $3$ \\ \hline
  ${\bf 40}, {\bf \bar {40}}$ & $2$ \\ \hline
  ${\bf 75}$& $0$ \\ \hline
\end{tabular}

\end{center}
\caption{Eigenvalues of the irreducible representations of
$\Lambda^6$ under the map $\Theta$. Since the map is real the
conjugates of the complex representations have the same
eigenvalues. }\label{table:maptheta}
\end{table}

So let us see what we can conclude from the above. {}First
consider the $\mathbf{75}$ part. This is projected out on the left
hand side of \p{thetaH} but is also not present on the right hand
side. So we have no contradiction here. Next consider the
$\mathbf{5}+\mathbf{\bar{5}}$ part. Again this is projected out by
$\Theta$ but is generically present on the left hand side. So we
conclude that the $(5,1)$ piece of $d(\Delta^{-3/2}\chi)$
vanishes. Equivalently, we conclude that the $(5,1)$ piece of
$d\chi$ corresponding to $W_5$  is exact: \be\label{dee} W_5=-12~d
\log\Delta \ee For the remaining representations the eigenvalues
are non-zero and \p{thetaH} allows us to determine the
corresponding $H_i$ in terms of the structure. We find,
\begin{eqnarray}\label{eqn:soln}
H_1&=&-\frac{3}{4} (d\omega)^{(0)}\Omega^2\nn
H_{10+\bar{10}}&=&-[\frac{1}{3}*(\Omega\wedge
d(\Delta^{-3/2}\chi))+(d\omega)^{(2,0)}+(d\omega)^{(0,2)}]\wedge
{\Omega}\nn
H_{24}&=&-\frac{1}{3}(d\omega)^{(1,1)}_0\wedge{\Omega}\nn
H_{40+\bar{40}}&=&\frac{1}{2}
*d(\Delta^{-3/2}\chi)-\frac{1}{6}*(\Omega\wedge
d(\Delta^{-3/2}\chi))\wedge\Omega
\end{eqnarray}
where $\Omega^n$ denotes the wedge product of $n$ factors of
$\Omega$.

At this point both $H_{75}$ and $H_{{5+\bar{5}}}$ are
undetermined. However, we can now use the equation for $dK$ in
\p{dk} to relate the $\bf{5+\bar{5}}$ part of $d\Omega$ and $H$ to
$\Delta$ giving: \bea\label{eqn:dD}
\frac{\Delta^{3/2}}{6~5!}*H_{mn_1...n_5}\chi^{n_1...n_5} =
\partial_m log\Delta-\frac{1}{6}\Omega^{r_1r_2}(d\Omega)_{m r_1 r_2}
\eea Looking at equations \p{eqn:dstarOmega} we find no further
constraints.

So let us summarize what we have learned. Any geometry admitting a
timelike Killing spinor can be written in the form
\bea\label{sumone} ds_{11}^2=-\Delta^2
(dt+\omega)^2+\Delta^{-1}g_{mn}dx^m dx^n \eea where the base space
with metric $g$ admits an $SU(5)$ structure $(g,\Omega,\chi)$
whose only restriction is that $W_5$ is exact and related to the
warp factor $\Delta$ via \be\label{sumtwo} W_5=-12d\log\Delta \ee
The four-form field strength can be written as:
\begin{eqnarray}\label{sumthree}F&=&
(dt+\omega)\wedge d\Omega
-[\frac{3}{4}(d\omega)^{(0)}\Omega+(d\omega)^{(2,0)}+(d\omega)^{(0,2)}
+\frac{1}{3}(d\omega)^{(1,1)}_0]\wedge\Omega\nn
&+&\frac{1}{2}*d(\Delta^{-3/2}\chi) -\frac{1}{2}*[\Omega\wedge
d(\Delta^{-3/2}\chi)]\wedge\Omega\nn &-&\frac{1}{16}
*([W_5+4W_4]\wedge \Delta^{-3/2}\chi) +F_{75}
\end{eqnarray}
where $F_{75}$ ($=H_{75}$) is an arbitrary 4-form on $B$ in the
$\mathbf{75}$ of $SU(5)$ (i.e. $F_{75} \in \Lambda^{(2,2)}_0$),
$W_4$ and $W_5$ are defined in \p{thedubi} and
$d\omega=d\omega^{(0)}\Omega+d\omega^{(1,1)}_0+
d\omega^{(2,0)}+d\omega^{(0,2)}$.

We started this section with the $SU(5)$ structure in D=11
specified by $K,\Omega,\Sigma$. However, our derivation of
\p{sumone}, \p{sumtwo}, \p{sumthree} mostly involved the $SU(5)$
structure in D=10 which is a component of the D=11 $SU(5)$
structure. The reason for this is that the D=11 structure is
constrained by the fact that $K$ is Killing, ${\cal
L}_K\Omega={\cal L}_K\Sigma=0$ and we worked with the obvious
adapted co-ordinates. Now $d\omega$ is an arbitrary
closed\footnote{Since $\omega$ need only be locally defined.}
two-form on $B$ as far as the D=10 SU(5) structure is concerned.
On the other hand, one can show that $dK$ specifies a part of the
intrinsic torsion of the D=11 structure, and since $dK=2d(log
\Delta)\wedge K-\Delta^2 d\omega$ we conclude that $d\omega$  is
in fact determined by the $D=11$ structure. This is in contrast to
$F_{75}$ which is determined by neither the D=10 nor the D=11
structure. See appendix E for further comments about the type of
$SU(5)$ structure in D=11.

We have thus derived necessary conditions on the local  form of
the most general bosonic geometry admitting a timelike Killing
spinor. The form includes a completely undetermined component of
the field strength and one might wonder if further conditions
might be imposed by considering the covariant derivatives of
$\Omega$ and $\Sigma$ and not just the exterior derivatives. To
see that they are not we will now prove a converse result: that
the above form of the metric and four-form field strength does
indeed admit a Killing spinor. In particular $F_{75}$ drops out of
the Clifford connection appearing in the Killing spinor equation.
It is for this reason that this component of the field-strength is
not determined by the Killing spinor equation alone.

To see this we first note that the geometry should preserve a
single Killing spinor $\epsilon$ giving rise to the $SU(5)$
structure in $D=11$. Such a spinor can be specified by demanding
that it be left inert by a number of projection operators. {}First
introduce the obvious orthonormal frame \bea e^0&=&\Delta(dt
+\omega)\nn e^i&=&\Delta^{-1/2}\bar e^i \eea where $\bar e^i$ is
an orthonormal frame for the base manifold $B$. The D=11 gamma
matrices give rise to D=10 gamma-matrices $\Gamma^i$ with
$\Gamma^0=\Gamma_{1\dots 10}$ proportional to the chirality
operator. It is convenient to introduce the chiral complex spinor
\be \eta=\frac{(1+ i \Gamma_0)}{\sqrt 2}\epsilon \ee in terms of
which \bea \Omega_{ij}&=&-i\eta^\dagger\Gamma_{ij}\eta\nn
\theta_{i_1\dots i_5}&=&i\eta^T\Gamma_{i_1\dots i_5}\eta \eea Both
the real an imaginary parts of $\eta$ give equivalent $SU(5)$
structures, but only the real part will be a Killing spinor, as we
shall see. Now the almost complex structure on the base manifold
is not integrable in general, and hence we cannot always introduce
complex co-ordinates. Nevertheless we can consistently introduce
holomorphic and anti-holomorphic tangent space indices which
simplifies the calculation. In terms of these we conclude that
$\eta$ satisfies the projections \be \Gamma^a\eta =0 \ee where
$a=1\dots 5$ is a holomorphic index and also \be \Gamma_{a_1\dots
a_5}\eta=-i\theta_{a_1\dots a_5}\eta^* \ee

We now consider the Killing spinor equation acting on
$\e=(\eta+\eta^*)/\sqrt 2$. Plugging in the expression for the
four-form and dealing with each $SU(5)$ irrep separately, we find
after a lengthy computation and using \p{spinids}, that
\begin{eqnarray}\label{finres}
[\nabla_{m}&+&\frac{1}{160}(\Omega W_5+5\Omega W_4)_m\Omega_{k_1
k_2}\Gamma^{k_1 k_2} -\frac{1}{16}(W_4)_k\Gamma_m{}^k \nn
&+&\frac{1}{8}\Omega_m{^r}(W_3)_{r k_1 k_2}\Gamma^{k_1 k_2}
-\frac{1}{394}\chi_{m k_1 k_2}{}^{n_1n_2}(W_1)_{n_1n_2}
\Gamma^{k_1 k_2}\nn &+&\frac{1}{192}\Omega_m{^r}(W_2)_{r
\ell_1\ell_2\ell_3}\chi^{\ell_1\ell_2\ell_3}{_{k_1
k_2}}\Gamma^{k_1 k_2}](\eta_0+\eta_0^*)=0
\end{eqnarray}
where we have rescaled the spinor $\eta\equiv \Delta^{1/2}\eta_0$
and used the notation $\Omega V_m\equiv\Omega_m{^r}V_r$. Now both
$\eta_0$ and  $\eta_0^*$ are solutions to this equation since the
connection is simply the sum of the Levi-Civita connection on $B$
with the intrinsic contorsion of the $D=10$ $SU(5)$ structure, as we show
in appendix C. However, one should not conclude that there are two
Killing spinors: the point is that \p{finres} only arises when the
Killing spinor equation is acting on the sum $(\eta+\eta^*)/\sqrt
2$ and not on $\eta,\eta^*$ separately. Thus we conclude that the
geometry in general preserves one Killing spinor
$\Delta^{1/2}(\eta_0+\eta_0^*)$ corresponding to just 1/32
supersymmetry.

Note that while the covariant derivative appearing in the original
Killing spinor equation of D=11 supergravity \p{killing} takes
values in the Clifford algebra, the covariant derivative appearing
in \p{finres} takes values in the spin sub-algebra. In other
words, the four-form field strength is necessarily constrained in
such a way that it transforms the Clifford connection into a spin
connection when acting on the preserved supersymmetries.

We have now shown that the form \p{sumone}, \p{sumtwo},
\p{sumthree} is both necessary and sufficient for a geometry to
admit a single time-like Killing spinor. However not all such
spacetimes are solutions of eleven dimensional supergravity. To
obtain solutions of the theory one just has to impose the gauge
equations of motion and the Bianchi identity for $F$ since the
Einstein equations will then be automatically be satisfied as we
showed in section 2.3. The Bianchi identity for $F$ can be
written 
\be\label{bone} d\omega\wedge d\Omega+dH=0 \ee while the
equation of motion for the four-form gives rise to two equations
\bea\label{btwo} d(\Delta^{-3}*d\Omega)+d\omega\wedge *H
+\frac{1}{2}H\wedge H&=&-\beta X_8\nn d\Omega\wedge H-d*H&=&0
\eea 
Here we have added in the correction term to the field
equation and have assumed that $i_K X_8=0$. Note that the third
equation is actually implied by the first (see the discussion
following \p{somewhat}). To check this in detail one can take the
exterior derivative of \p{thetaH} to find \be
d(*\Theta(H))-d\omega\wedge d\Omega\wedge \Omega=0 \ee and then
substitute \p{bone}. Note that one can further substitute the
expression for $H$ given by \p{eqn:soln} and \p{eqn:dD} into
\p{bone}, \p{btwo} but as the result is not too illuminating we
shall not present it here. It is worth emphasising that the
component of the four-form not determined by the Killing spinor
equation, $F_{75}$, is constrained and related to the $SU(5)$
structure by the Bianchi identity and the equations of motion

At this stage we can present some vanishing theorems when the
ten-dimensional base manifold $B$ is compact. Consider first the
case when $H=0$ and hence $F=(dt+\omega)\wedge d\Omega$. We then
have \be \int_B \Delta^{-3}d\Omega\wedge *d\Omega= -\int_B
\Omega\wedge d(*\Delta^{-3}d\Omega)=0 \ee where we have integrated
by parts and then used the equation of motion \p{btwo} (ignoring
the $X_8$ correction). Since the left hand side of the equation is
positive semi-definite we conclude that $d\Omega=0$ which in turn
implies that the four-form $F=0$.

Let us now obtain two results for non-vanishing $H$. First observe
that using \p{thetaH} and \p{eqn:soln} we deduce
\be\label{blue}
d(\Delta^{-3/2}\chi)=*\Theta(H)-\frac{1}{2}d\omega\wedge\Omega^2
=*\Theta(H') \ee where $H'$ is defined to be the pieces of $H$
that are independent of $d\omega$. Next consider \be \int_B
*\Theta(H')\wedge \Theta(H')=-\int_B \Delta^{-3/2}\chi\wedge d
\Theta(H') \ee where we have integrated by parts. We next note
that \be\label{exp} d \Theta(H')=d(-H'_{10+\bar{10}}
+2H_{40+\bar{40}}) \ee On the other hand we know from the Bianchi
identity \p{bone} that \be d\omega\wedge d\Omega+dH=0 \ee If we
now restrict to $d\omega=d\omega^{(2,0)}+d\omega^{(0,2)}$ then
this equation becomes \be\label{ex} d(H_{5+\bar
5}-H'_{10+\bar{10}} +H_{40+\bar{40}}+H_{75})=0 \ee Comparing with
\p{exp}  suggests we should further restrict to $H_{5+\bar
5}=H_{75}=0$.

We now obtain two theorems. If
$d\omega=d\omega^{(2,0)}+d\omega^{(0,2)}$, $H_{5+\bar 5}=H_{75}=0$
and $H_{40+\bar{40}}=0$ then \p{ex}, \p{exp} and \p{blue} implies
that $H_{10+\bar{10}}=0$ also and hence $H$=0. Similarly if
$d\omega=d\omega^{(2,0)}+d\omega^{(0,2)}$, $H_{5+\bar 5}=H_{75}=0$
and $H_{10+\bar{10}}=0$ then $H_{40+\bar{40}}=0$ also and hence
$H$=0. In both cases the previous result assuming $H=0$ and
arbitrary $d\omega$ then implies that $F=0$.

\section{Examples of solutions with $SU(5)$ structures}

In order to gain some further insight into the formalism, we will
now display $SU(5)$ structures for some known solutions. As a
bonus, in carrying out this exercise we will be able to spot some
new solutions.

\subsection{M5 branes}

Let us first look at the simple $M5$-brane solution. The metric
and field strength can be written as:
\begin{eqnarray}\label{eqn:M5}
ds_{11}^2&=&H^{-1/3}(-dt^2+dx^i dx^i)+H^{2/3}dy^idy^i
\nn \ast_{11}F&=&-dt\wedge dx^1\wedge dx^2\wedge dx^3\wedge
dx^4\wedge dx^5 \wedge dH^{-1}
\end{eqnarray}
with $i=1,\dots,5$ and $H=H(y)$ a harmonic function. This solution
is well known to preserve 1/2 of the supersymmetry: the Killing
spinors satisfy the single projection $\Gamma_{012345}\e=\e$.
There are certainly timelike spinors which satisfy this projection
and so we should be able to display a $SU(5)$ structure for it.

Comparing with \p{sumone} we identify $\Delta=H^{-1/6}$ and the
base space metric is then given by,
\begin{equation}\label{eqn:baseM5}
ds^2=H^{-1/2}dx^i dx^i+H^{1/2}dy^i dy^i
\end{equation}
Define the complex $(1,0)$ frame,
\begin{equation}\label{eqn:structure}
\Theta^i=H^{-1/4}dx^i+i H^{1/4}dy^i
\end{equation}
The corresponding $SU(5)$ structure is given by:
\begin{eqnarray}\label{eqn:M5structure}
\Omega&=&\frac{i}{2}\Theta^{i}\wedge \bar{\Theta}^i \nn
\chi&=&Re(\Theta^1\wedge\Theta^2\wedge\Theta^3\wedge\Theta^4\wedge\Theta^5)
\end{eqnarray}
In terms of the coordinate basis these are given by
\begin{eqnarray}\label{eqn:M5structure-coordinate}
\Omega&=&dx^i\wedge dy^i \nn
\chi&=&H^{-5/4}dx^{12345}-\frac{H^{-1/4}}{3!
2!}\varepsilon_{i_1....i_5}dx^{i_1i_2i_3}\wedge dy^{i_4i_5}
+\frac{H^{3/4}}{4!}\varepsilon_{i_1...i_5}dx^{i_1}\wedge
dy^{i_2...i_5}
\end{eqnarray}
where $\varepsilon_{i_1...i_5}$ is just the $d=5$ permutation
symbol.

To see that this $SU(5)$ structure is indeed related to a Killing
spinor we first note, after a small calculation, that
$d(\Delta^{-3/2}\chi)$ has no $(5,1)+(1,5)$ pieces and hence
\p{sumtwo} is satisfied. This is the only restriction required on
the structure, but we note that here we also have $d\Omega=0$. We
next need to show that the four-form can be recovered from
\p{sumthree}. Interestingly, to achieve this it is necessary to
include a non-vanishing $F_{75}$. Specifically we set
\begin{equation}
\label{eqn:M5F75}F_{75}= \frac{1}{16\cdot 2!2!}\partial^{i}log
H\varepsilon_{i j_1...j_4}\Theta^{j_1}\wedge
\Theta^{j_2}\wedge\bar{\Theta}^{j_3}\wedge\bar{\Theta}^{j_4}
\end{equation}
and then \p{sumthree} agrees with the expression in \p{eqn:M5}.

Since the fivebrane solution preserves 16 Killing spinors the
solution has more than one $SU(5)$ structure. Note also that some
of the Killing spinors can be null so that the solution also
belongs to the null class. It would be interesting to display the
$SU(5)$ structure for the solution corresponding to a fivebrane
wrapping a SLAG five-cycle \cite{Gauntlett:2000ng}, as this
solution preserves just 1/32 supersymmetry .

\subsection{Flat and resolved $M2$ branes}

Let us now recover some known solutions involving membranes. We
take the ten-dimensional base space $B$ to be of the form:
\begin{equation}\label{ch6:M2metric}
ds^2=\Delta^3(dx_1^2+dx_2^2)+ds^2(M_8)
\end{equation}
where  $ds^2(M_8)$ is a Ricci flat metric with holonomy contained
in $SU(4)$. One can then define the following $SU(5)$ structure
\begin{eqnarray}\label{ch6:M2structure}
\Omega&=&\Delta^3 dx^1\wedge dx^2+\omega_{(8)}\nonumber\\
\chi&=&\Delta^{3/2}(dx^1\wedge \hat{\chi}_1+dx^2 \hat{\chi}_2)
\end{eqnarray}
where we have introduced the K\"ahler form $\omega_{(8)}$ and
holomorphic $(4,0)$ form $\hat{\theta}=\hat{\chi}_1-i
\hat{\chi}_2$ of $M_8$. The base space has in fact an
$SU(4)\subset SU(5)$ structure. Note that the normalizations of
$\Omega$ and $\chi$ are not arbitrary but are chosen to ensure
that $\Omega$ is a K\"ahler form for the base space and that they
satisfy the compatibility condition \p{ch6:compatibility}. Also we
could have chosen an arbitrary function $f^2$ in the metric
instead of $\Delta^3$ but demanding that \p{sumtwo} is satisfied
implies that $f^2=\Delta^3$.

{}For simplicity we will assume that $\Delta$ does not depend on
$(x^1,x^2)$. This implies that the ${\bf 5}+{\bf \bar 5}$ piece of
the spatial part of the four-form field strength vanishes. Let us
first restrict to static solutions and set the rotation parameter
to zero. The expression for the four-form field strength
\p{sumthree} becomes
\begin{equation}
\label{ch6:M2flux}F=dt\wedge dx^1\wedge dx^2\wedge
d\Delta^3+F_{75}
\end{equation}
where $F_{75}$ is any four form on the base space in the
\textbf{75} of $SU(5)$. Imposing the Bianchi identity and gauge
equations of motion and using $*F_{75}=F_{75}\wedge \Omega$~one
finds:
\begin{eqnarray}\label{eqn:M2gauge}
d F_{75}&=&0 \nonumber\\
d*_8 d\Delta^{-3}-\frac{1}{2}F_{75}\wedge F_{75}&=&-\beta X_8
\end{eqnarray}
where $*_8$ is the Hodge star with respect to the 8 dimensional
metric.

In the simple case that $F_{75}=0$ and flat transverse space, we
recover the well known 1/2 supersymmetric $M2$ brane solution
\begin{eqnarray}\label{ch6:M2SU(4)}
ds^2&=&H^{-2/3}(-dt^2+dx_1^2+dx_2^2)+H^{1/3}ds^2(\bE^8) \nonumber
\\F&=&dt\wedge dx^1\wedge dx^2\wedge d(H^{-1})
\end{eqnarray}
where $H\equiv \Delta^{-3}$ is harmonic.

Another possibility is to take $F_{75}$ to be a four form on
$M_8$. Under $SU(4)\subset SU(5)$ we have the following
decomposition:
$\mathbf{75}\rightarrow\mathbf{15}+\mathbf{20}+\mathbf{\bar{20}}+\mathbf{20'}$.
The first three representations occur when one of the indices of
$F_{75}$ is in the $(x^1,x^2)$ directions so for $F_{75}$ to be a
four form on $M_{8}$ it must belong to the $\mathbf{20'}$ of
$SU(4)$ i.e. it must be a self-dual $(2,2)$ form. Since it must be
closed it follows that $F_{75}=L_{(2,2)}$ with $L_{(2,2)}$ a
harmonic self dual four form. This modifies the equation for $H$
and we get
\begin{equation}\label{ch6:resolvedH}
\square H=-\frac{1}{2}|L_{(2,2)}|^2 +\beta  X_8
\end{equation}
Thus we recover the resolved 1/8 supersymmetric $M2$ brane
solutions of
\cite{Becker:1996gj,Duff:1997hf,Hawking:1997bg,Cvetic:2000mh,Cvetic:2000db}.
As for the fivebrane solution, these solutions also belong to the
null class.

A simple rotating generalisation of these solutions is to choose
$d\omega$ to be the sum of a $(2,0)+(0,2)$ form. Specifically,
given a closed two-form $\alpha\in \Lambda^{2,0}( M_{8})$ we set
$d\omega=\alpha+\bar{\alpha}$ and get the supersymmetric solution
\begin{eqnarray}
ds^2&=&H^{-2/3}[-(dt+\omega)^2+dx_1^2+dx_2^2]+H^{1/3}ds^2(M_8)
\nonumber \\F&=&(dt+\omega)\wedge dx^1\wedge dx^2\wedge d(H^{-1})
+F_{75}
\end{eqnarray}
with \p{eqn:M2gauge} unchanged.

\subsection{Rotating Calabi-Yau and the G\"{o}del solution}

Recently, an interesting generalisation of the G\"{o}del solution
was found in five-dimensions \cite{Gauntlett:2002nw}. Uplifted to
D=11 it was shown to preserve 5/8 of the supersymmetry. We now
show that this fits into a broader class of new solutions.

We look for rotating solutions with no warp factor for which the
base space $M_{10}$ is a complex manifold with holonomy
$G\subseteq SU(5)$. Then, the only non-zero components of the
field strength can arise from the rotation and from $F_{75}$.
Similarly to the last sub-section, we set
$d\omega=\alpha+\bar{\alpha}$ with $\alpha\in \Lambda^{2,0}(
M_{10})$. We then find the supersymmetric solution
\begin{eqnarray}\label{eqn:rotCY}
ds_{11}^2&=&-(dt+\omega)^2+ds^2(M_{10})\nn F&=&-d\omega\wedge
\Omega  +F_{75}
\end{eqnarray}
provided that $dF_{75}=0$ and $F_{75}\wedge F_{75}=\beta X_8$.

As a particular example of this class of solutions we take the
base space to be flat $\bE^{10}$ and set $F_{75}=0$. Introduce
complex coordinates $z^a$  for $\bE^{10}$ and the canonical
$SU(5)$ structure,
\begin{eqnarray}\label{eqn:flatSU(5)}  \Omega&=&\frac{i}{2}dz^a\wedge
d\bar{z}^a \nn  \chi&=&Re(dz^1\wedge dz^2\wedge dz^3\wedge
dz^4\wedge dz^5)
\end{eqnarray}
One can then choose $\alpha=dz^1\wedge dz^2$ and this gives the
G\"{o}del solution of \cite{Gauntlett:2002nw} which preserves 5/8
supersymmetry. Further G\"odel-type solutions are obtained by
choosing different $\alpha\in \Lambda^{2,0}(\bE^{10})$, and it
would be interesting to see which of them preserve exotic
fractions of supersymmetry.

\section{Conclusion}

We have shown that the most general supersymmetric configurations
of D=11 supergravity, preserving at least one Killing spinor, have
either a privileged $SU(5)$ or a $(Spin(7)\ltimes\bR^8)\times \bR$
structure constructed from the Killing spinor, depending on
whether the vector constructed from a bi-linear of the Killing
spinor is time-like or null, respectively. {}For the time-like
case, we carried out a detailed local analysis using the Killing
spinor equation: we found that the $SU(5)$ structure in D=11 is
restricted by the fact that the time-like vector is always Killing
and that the $SU(5)$ structure of the D=10 base space orthogonal
to the orbits of the Killing vector is only weakly constrained. We
deduced the necessary and sufficient conditions on the form of the
geometry admitting Killing spinors and showed that most of the
form is determined by the D=11 $SU(5)$ structure. In particular
there was a component of the four-form which dropped out of the
Killing spinor equation and hence is undetermined. We also
analysed what extra constraints are imposed in order to ensure
that the geometries preserving Killing spinors also solve the
equations of motion. These constraints relate the component of the
four-form undetermined by the Killing spinor equation to the
$SU(5)$ structure. To complete the classification of the most
general supersymmetric solutions we need to carry out a similar
analysis for the null case.

The formalism was used to prove some no-go theorems when the D=10
base space is compact and also to construct some new solutions. It
would be interesting to study the new solutions further. For
example it would be interesting to determine the fraction of
supersymmetry preserved by the new G\"odel-type solutions
presented at the end of section 5.3.

We have also proposed a finer classification for configurations
that preserve more than one supersymmetry. Such configurations
will have various numbers of different $SU(5)$ and/or
$(Spin(7)\ltimes\bR^8)\times \bR$ structures that can be
constructed from the Killing spinors, or equivalently, a
privileged $G$-structure with $G\subset SU(5)$ or $G\subset
(Spin(7)\ltimes\bR^8)\times \bR$. The first step then, is to
classify these $G$-structures which are defined to be the isotropy
groups of the Killing spinors. They can also be specified by
algebraic conditions on the tensors that can be constructed from
bi-linears in the spinors. The second step in the classification
is to then use the Killing spinor equation to place constraints on
the $G$-structure, as well as to solve for various parts of the
metric and four-form field strength in terms of the structure. It
would be quite an achievement to carry out this programme in full.

\section*{Acknowledgements}
We would like to thank Chris Hull for collaboration in the early
stages of this work and, with Daniel Waldram, for very helpful
discussions. We would also like to thank Jos\'e Figueroa-O'Farrill
for useful comments on an earlier draft.

\appendix
\makeatletter
\renewcommand{\theequation}{A.\arabic{equation}}
\@addtoreset{equation}{section} \makeatother
\section{Conventions}

We use the signature $(-,+,...,+)$. D=11 co-ordinate indices will
be denoted $\mu,\nu,\dots$ while tangent space indices will be
denoted by $\alpha,\beta,\dots$.

The D=11 spinors we will use are Majorana. The gamma matrices
satisfy \be \{\Gamma_\alpha,\Gamma_\beta\}=2\eta_{\alpha\beta} \ee
and can be taken to be real in the Majorana representation. They
satisfy, in our conventions, $\Gamma_{012345678910}=1$ and as a
consequence of this the following duality relation holds: \be
\Gamma_{\alpha_1\alpha_2...\alpha_p}=(-1)^{\frac{(p+1)(p-2)}{2}}\frac{1}{(11-p)!}
\varepsilon{_{\alpha_1\alpha_2...\a_p}}{^{\a_{p+1}\a_{p+2}...\a_{11}}}
\Gamma_{\a_{p+1}\a_{p+2}...\a_{11}} \ee where we have defined \be
\varepsilon_{012345678910}=+1 \ee It follows that
$\Delta=\basismu$ is a basis of the Clifford algebra
$C\ell(10,1)\cong R(32)$, where $R(32)$ is the algebra of
$32\times 32$ matrices. We will use repeatedly the following
formula for anti-symmetrising products of gamma matrices:
\be\Gamma_{\a_1\a_2...\a_n}
\Gamma^{\b_1\b_2...\b_m}=\sum_{k=0}^{k=min(n,m)}
{\frac{m!n!}{(m-k)!(n-k)!k!}}
\Gamma{_{[\a_1\a_2...\a_{n-k}}}{^{[\b_{k+1}
...\b_m}}\delta^{\b_1}_{\a_n}\delta^{\b_2}_{\a_{n-1}}...
\delta^{\b_{k}]}_{\a_{n-k+1}]} \ee For any $M,N$ $\in$ $R(32)$ we
can perform a Fierz rearrangement using:
\begin{eqnarray}
M{_{a}}{^b}N{_c}{^c} &=& \frac{1}{32}\{(NM){_a}{^d}\delta{_c}{^b}
+(N\Gamma^{\a_1}M){_a}{^d}(\Gamma_{\a_1}){_c}{^b} \nn
&-&\frac{1}{2!}(N\Gamma^{\a_1\a_2}M){_a}{^d}(\Gamma_{\a_1\a_2}){_c}{^b}
-\frac{1}{3!}(N\Gamma^{\a_1\a_2\a_3}M){_a}{^d}(\Gamma_{\a_1\a_2\a_3}){_c}
{^b} \nn
&+&\frac{1}{4!}(N\Gamma^{\a_1\a_2\a_3\a_4}M){_a}{^d}(\Gamma_{\a_1\a_2\a_3\a_4}
){_c}{^b}
+\frac{1}{5!}(N\Gamma^{\a_1\a_2\a_3\a_4\a_5}M){_a}{^d}(\Gamma_{\a_1\a_2\a_3
\a_4\a_5}){_c}{^b}\} \nn
\end{eqnarray}
where $a,b,c,d=1,\dots,32$.

\par Given a Majorana spinor $\e$ its conjugate  is given by $\bar{\e} =\e^T
C$, where $C$ is the charge conjugation matrix in D=11 and
satisfies $C^T=-C$. In the Majorana representation we can choose
$C=\Gamma_0$. An important property of gamma matrices in D=11 is
that the matrix $C\Gamma_{\a_1\a_2...\a_p}$ is symmetric for
$p=1,2,5$ and antisymmetric for $p=0,3,4$ (the cases $p>5$ are
related by duality to the above). For an antisymmetrized product
$\Gamma_{(n)}$ of $n$ gamma matrices and any spinor $\e$ we have :
\begin{equation}\label{gammabar}
   \overline{\Gamma_{(n)}
\e}=(-1)^{\frac{n(n+1)}{2}}\bar{\e}\Gamma_{(n)}
\end{equation}

The Hodge star of a $p$-form $\omega$ is defined by \be
*\omega_{\mu_1\dots \mu_{11-p}}=\frac{\sqrt
{-g}}{p!}\epsilon_{\mu_1\dots \mu_{11-p}}
{}^{\nu_1\dots\nu_p}\omega_{{\nu_1\dots\nu_p}} \ee and the square
of a $p$-form via \be
\omega^2=\frac{1}{p!}\omega_{\mu_1\dots\mu_p}\omega^{\mu_1\dots\mu_p}
\ee

\section{Integrability conditions from the Killing spinor equation}
\makeatletter
\renewcommand{\theequation}{B.\arabic{equation}}
\@addtoreset{equation}{section} \makeatother

Taking the second covariant derivative of the Killing spinor
equation \p{killing} and anti-symmetrising we obtain: \bea
\nabla_{[\rho}\nabla_{\mu]}\e&=&-\frac{1}{288}(\Gamma{_{[\mu}}{^{\nu_1\nu_2\nu_3\nu_4}}-8\delta_{[\mu}^{\nu_1}\Gamma^{\nu_2\nu_3\nu_4})
\nabla_{\rho]}F_{\nu_1\nu_2\nu_3\nu_4}\e \nn
&+&\frac{1}{288^2}(\Gamma{_{[\mu}}{^{\nu_1\nu_2\nu_3\nu_4}}-8\delta_{[\mu}^{\nu_1}\Gamma^{\nu_2\nu_3\nu_4})(\Gamma{_{\rho]}}{^{\s_1\s_2\s_3\s_4}}-8\delta_{[\rho}^{\s_1}\Gamma^{\s_2\s_3\s_4})F_{\nu_1\nu_2\nu_3
\nu_4}F_{\s_1\s_2\s_3\s_4}\e\nn \eea The terms on the right hand
side can be simplified using the identity: \bea
&(\Gamma{_{[\mu}}{^{\nu_1\nu_2\nu_3\nu_4}}-8\delta_{[\mu}^{\nu_1}\Gamma^{\nu_2\nu_3\nu_4})
(\Gamma{_{\rho]}}{^{\s_1\s_2\s_3\s_4}}-8\delta_{\rho]}^{\s_1}\Gamma^{\s_2\s_3\s_4})=\nn
&\Gamma{_{\mu\rho}}{^{\nu_1\nu_2\nu_3\nu_4\s_1\s_2\s_3\s_4}}+16
\delta_{[\mu}^{\nu_1}\Gamma{_{\rho]}}{^{\nu_2\nu_3\nu_4\s_1\s_2\s_3\s_4}}+2\cdot
4!\delta_{\mu\rho}^{\nu_1\s_1}\Gamma^{\nu_2\nu_3\nu_4\s_2\s_3\s_4}\nn
&+4 \cdot 4!
\delta_{[\mu}^{\nu_1}g^{\nu_2\s_1}\Gamma{_{\rho]}}{^{\nu_3\nu_4\s_2\s_3\s_4}}
-3\cdot 4!
g^{\nu_1\s_1}g^{\nu_2\s_2}\Gamma{_{\mu\rho}}{^{\nu_3\nu_4\s_3\s_4}}
-4!4!\delta_{[\mu}^{\nu_1}g^{\nu_2\s_1}g^{\nu_3\s_2}\Gamma_{\nu]}{^{\nu_4\s_3\s_4}}\nn
&+16\cdot 4!
\delta_{\mu\rho}^{\nu_1\nu_2}g^{\nu_3\s_1}\Gamma^{\nu_4
\s_2\s_3\s_4} +4! g^{\nu_1\s_1} g^{\nu_2\s_2} g^{\nu_3\s_3}
g^{\nu_4\s_4}\Gamma_{\mu\rho} -8\cdot 4! \delta_{[\mu}^{\nu_1}
g^{\nu_2\s_1} g^{\nu_3\s_2}
g^{\nu_4\s_3}\Gamma{_{\rho]}}{^{\s_4}}\nn &-36\cdot 4!
\delta_{\mu\rho}^{\nu_1\s_1} g^{\nu_2\s_2}
g^{\nu_3\s_3}\Gamma^{\nu_4\s_4} \qquad (AS) \eea where ``AS''
refers to the fact that this equation is true when  we
anti-symmetrise over the indices $\s_1,\dots,\s_4$ and
$\nu_1,\dots,\nu_4$. Also, the left hand side can be expressed in
terms of the Riemann tensor via \bea
\nabla_{[\rho}\nabla_{\mu]}\e=\frac{1}{8}R_{\rho\mu\s_1\s_2}\Gamma^{\s_1\s_2}\e
\eea The resulting integrability condition first appeared in
\cite{Biran:1982eg}.

We note that the integrability condition can be recast in the form
\begin{eqnarray}
\big( \sum_{r=1}^{r=5}
\frac{1}{r!}L^{(r)}_{\mu\nu}{^{\s_1..\s_r}}\Gamma_{\s_1..\s_r}\big)\e=0
\end{eqnarray}
where we have,
\begin{eqnarray}
L^{(1)}_{\mu\nu}{^{\s_1}}&=&-\frac{1}{144}*(F\wedge
F)_{\mu\nu}{^{\s_1}}\nn
L^{(2)}_{\mu\nu}{^{\s_1\s_2}}&=&-\frac{1}{4}R_{\mu\nu}{^{\s_1\s_2}}-\frac{1}{72}F^2
\delta_{[\mu}^{\s_1}\delta_{\nu]}^{\s_2}
-\frac{1}{216}\delta_{[\mu}^{\s_1}F_{\nu]}{^{\l_1\l_2\l_3}}
F_{\l_1\l_2\l_3}{^{\s_2}}\nn
&~~&+\frac{1}{48}F_{[\mu}{^{\s_1\l_1\l_2}}F_{\nu]\l_1\l_2}{^{\s_2}}\nn
L^{(3)}_{\mu\nu}{^{\s_1\s_2\s_3}}&=&\frac{1}{6}\nabla_{[\mu}F_{\nu]}{^{\s_1\s_2\s_3}}
+\frac{1}{216}*F_{[\mu}{^{\s_1\s_2\s_3\l_1\l_2\l_3}}F_{\nu]\l_1\l_2\l_3}\nn
L^{(4)}_{\mu\nu}{^{\s_1\s_2\s_3\s_4}}&=&\frac{1}{9}F_{\mu\nu}{^{\s_1\l_1}}F_{\l_1}{^{\s_2\s_3\s_4}}
-\frac{1}{6}\delta_{[\mu}^{\s_1}F_{\nu]}{^{\s_2\l_1\l_2}}F_{\l_1\l_2}{^{\s_3\s_4}}\nn
L^{(5)}_{\mu\nu}{^{\s_1\s_2\s_3\s_4\s_5}}&=&\frac{5}{12}\delta_{[\mu}^{\s_1}\nabla_{\nu]}F^{\s_2\s_3\s_4\s_5}
-\frac{1}{1728}\e_{\l_1\l_2\l_3\l_4\l_5\l_6}{^{\s_1\s_2\s_3\s_4\s_5}}F_{[\mu}{^{\l_1\l_2\l_3}}
F_{\nu]}{^{\l_4\l_5\l_6}}\nn
&~~&-\frac{1}{864}\e_{\l_1\l_2\l_3\l_4\l_5[\mu}{^{\s_1\s_2\s_3\s_4\s_5}}F_{\nu]}{^{\l_1\l_2\kappa_1}}
F_{\kappa_1}{^{\l_3\l_4\l_5}}\nn
&~~&+\frac{1}{1152}\e_{\mu\nu\l_1\l_2\l_3\l_4}{^{\s_1\s_2\s_3\s_4\s_5}}
F^{\l_1\l_2\kappa_1\kappa_2}F_{\kappa_1\kappa_2}{^{\l_3\l_4}}
\end{eqnarray}
If we now contract with $\Gamma^{\mu}$ and use the Bianchi
identity $R_{\mu[\nu\rho\s]}=0$ we find that:
\bea 0&=&[R_{\rho\mu}-\frac{1}{12}(F_{\rho \s_1\s_2\s_3}F{_{\mu}}
{^{\s_1\s_2\s_3}}-\frac{1}{12}g_{\rho\mu}F^2)]\Gamma^{\mu}\e\nn
&-&\frac{1}{6\cdot 3!}*(d*F+\frac{1}{2}F \wedge
F)_{\s_1\s_2\s_3}(\Gamma_{\rho}{^{\s_1\s_2\s_3}}-6\delta^{\s_1}_{\rho}\Gamma^{\s_2\s_3})\e\nn
&-&\frac{1}{6!}dF_{\s_1\s_2\s_3\s_4\s_5}(\Gamma_{\rho}{^{\s_1\s_2\s_3\s_4\s_5}}
-10\delta_{\rho}^{\s_1}\Gamma^{\s_2\s_3\s_4\s_5})\e \eea

\section{$SU(5)$ structures in ten dimensions}
\makeatletter
\renewcommand{\theequation}{C.\arabic{equation}}
\@addtoreset{equation}{section} \makeatother

Consider a $SU(5)\subset SO(10)$ structure specified by
$g,\Omega,\chi$ or equivalently by a chiral spinor $\eta$. As
mentioned earlier, there always exists a connection $\nabla'$ that
preserves the structure, $\nabla'\eta=0$. In fact it is not unique
and there is a whole family of such connections. The intrinsic
torsion of the $SU(5)$ structure is the part of the torsion of an
$SU(5)$ preserving connection that does not depend on the specific
choice of such a connection. It can thus be thought of as an
equivalence class of torsion tensors. Let us explain this in more
detail.

Any metric preserving connection can be written as
$\Gamma{^{r}}{_{mn}}=C{^r}{_{mn}}+K{^r}{_{mn}}$, where
$C{^r}{_{mn}}$ are the Christoffel symbols and $K^r{_{mn}}$ is
called the contorsion tensor. The contorsion satisfies the
symmetry property $K_{rmn}=-K_{nmr}$ and the torsion can be
determined by the contorsion by $T^r{_{mn}}=2K^r{_{[mn]}}$. One
can also construct the contorsion tensor from the torsion as
\begin{equation}
K^r{_{mn}}=\frac{1}{2}(T^r{_{mn}}+T_m{^{r}}{_n}+T_n{^{r}}{_m})
\end{equation}
Thus the torsion and contorsion are essentially equivalent.

The contorsion (and torsion) tensor lives in the space $T^*\otimes
so(10)\simeq (T^*\otimes su(5))\oplus(T^*\otimes su(5)^{\perp})$,
where $su(5)^{\perp}$ is the orthogonal complement of $su(5)$ in
$so(10)$. The part of the contorsion tensor that lies in
$T^*\otimes su(5)$ acts trivially on $SU(5)$ singlets such as
$\Omega$ and $\chi$. Thus any two connections that preserve the
$SU(5)$ structure will differ by an element of $T^*\otimes su(5)$
and so the intrinsic contorsion, which we denote $K^0$, is the
part of the contorsion that lies in $T^*\otimes su(5)^{\perp}$.

The space $T^*\otimes su(5)^{\perp}$, where the intrinsic
contorsion lies, decomposes as \be ({\bf 5}+{\bf \bar{5}})\times
({\bf 1}+{\bf 10}+{\bf \bar{10}})\to ({\bf 10}+{\bf \bar{10}})+
({\bf 40}+{\bf \bar{40}})+ ({\bf 45}+{\bf \bar{45}})+ ({\bf
5}+{\bf\bar{5}})+({\bf 5'}+{\bf \bar{5}'}) \ee while the space
$T^*\otimes su(5)$ decomposes as,
\begin{equation}\label{Ttimesg}
({\bf 5}+{\bf \bar{5}})\times ({\bf 24})\to ({\bf 5}+{\bf
\bar{5}})+({\bf 45}+{\bf \bar{45}})+({\bf 70}+{\bf \bar{70}})
\end{equation}
We thus see that the contorsion tensor has, generically, three
${\bf 5}+{\bf \bar{5}}$ and two ${\bf 45}+{\bf \bar{45}}$ pieces.
The most general contorsion tensor can thus be written
\begin{eqnarray}\label{contorsion}
K_{rmn}&=&T^{(1)}{_m}
\Omega_{nr}+T^{(2)}{_{[n}}\Omega_{r]m}+T^{(3)}{_{[n}}g_{r]m}\\\nonumber
&+&(P^{2,1}_0{_{rmn}}+\Omega{_m}{^k}Q^{2,1}_0{_{k
nr}}+c.c.)\\\nonumber &+&(S^{3,1}_0{_{m
k_1k_2k_3}}\chi^{k_1k_2k_3}{_{nr}}+c.c.)\nn
&+&(R^{3,0}_{mnr}+c.c.)+K_{rmn}^{70+\bar{70}}
\end{eqnarray}
where $c.c.$ stands for complex conjugate, $P,Q,R,S$ are forms of
the type indicated. Demanding that the above connection preserves
the $SU(5)$ structure allows one to relate the various components
of the contorsion tensor to the $W_i$ defined in \p{thedubi}.

To do this let $\nabla'$ be a covariant derivative with contorsion
$K$ that preserves the $SU(5)$ structure. To proceed write
$\nabla' \Omega=\nabla'\chi=0$ and then anti-symmetrise all of the
indices to get \bea \frac{1}{6}d\Omega_{n_1n_2n_3}&=&K^{r}{_{[n_1
n_2}}\Omega{_{|r| n_3]}}\nn
\frac{1}{30}d\chi_{n_1...n_6}&=&K^{r}{_{[n_1n_2}}\chi{_{|r|n_3 n_4
n_5 n_6]}} \eea It is now possible to explicitly relate the irreps
of $K$ to $W_i$ by decomposing the left and right hand sides into
$SU(5)$ irreps. We find that $R$ and $S$ are uniquely determined
by $W_1$ and $W_2$, respectively, and that
\begin{eqnarray}\label{contfreedom}
(W_3){_{n_1n_2n_3}}&=&-6\Omega_{[n_1}{^r}P^{2,1}_0{_{n_2n_3]
r}}-2Q^{2,1}_0{_{n_1n_2n_3}}+c.c.\nn
(W_4){_m}&=&-4\Omega_m{^r}T^{(2)}_r-4 T^{(3)}_m \nn
(W_5){_m}&=&40\Omega_m{^r}T^{(1)}_r+12\Omega_m{^r}T^{(2)}_r+20
T^{(3)}
\end{eqnarray}
The fact that the above components of the contorsion are not
uniquely determined in terms of the $W_i$ reflects the freedom in
defining an $SU(5)$ preserving connection. Solving for
$T^{(1)},T^{(2)}$ in terms of $W_4, W_5, T^{(3)}$ and for
$P^{2,1}_0$ in terms of $W_3, Q^{2,1}_0$ we conclude that the
contorsion can be expressed as
\begin{eqnarray}\label{contcomp}
K_{rmn}&=&-\frac{1}{40}\big( \Omega W_5+3 \Omega
W_4\big)_m\Omega_{nr}+\frac{1}{4}(\Omega W_4)_{[n}\Omega_{r
]m}+\frac{3}{2}\Omega_{[r}{^k}(W_3)_{mn]k}\nn &+&\frac{1}{4\cdot
4!}\chi_{rmn}{^{k_1k_2}}(W_1)_{k_1 k_2}-\frac{1}{2\cdot
4!}\Omega_m{^k}(W_2)_{k\ell_1\ell_2\ell_3}\chi{^{\ell_1\ell_2\ell_3}}{_{nr}}\nn
&+&\Big(\frac{1}{5}\Omega T^{(3)}_m \Omega_{nr}+\Omega
T^{(3)}{_{[n}}\Omega_{r]m}+T^{(3)}{_{[n}}g_{r]m}\Big)\nn
&+&\Big(3\Omega{_{[r}}{^k}Q^{2,1}_{0}{_{mn]k}}
+\Omega{_{m}}{^k}Q^{2,1}_{0}{_{nrk}}+c.c.\Big)+K_{rmn}^{70+\bar{70}}
\end{eqnarray}
where we have used the notation $\Omega V_m\equiv\Omega_m{^r}V_r$.
The last three terms in the brackets act trivially on $\Omega$ and
$\chi$ and so correspond to the terms appearing in the
decomposition \p{Ttimesg}. Thus the intrinsic contorsion can be
defined by \p{contcomp} with the last three terms set to zero. We
thus have
\begin{eqnarray}\label{noname}
K^0_{rmn}&=&-\frac{1}{40}\big( \Omega W_5+3 \Omega
W_4\big)_m\Omega_{nr}+\frac{1}{4}(\Omega W_4)_{[n}\Omega_{r
]m}+\frac{3}{2}\Omega_{[r}{^k}(W_3)_{mn]k}\nn &+&\frac{1}{4\cdot
4!}\chi_{rmn}{^{k_1k_2}}(W_1)_{k_1 k_2}-\frac{1}{2\cdot
4!}\Omega_m{^k}(W_2)_{k\ell_1\ell_2\ell_3}\chi{^{\ell_1\ell_2\ell_3}}{_{nr}}
\end{eqnarray}
Equivalently we can calculate from this the intrinsic torsion thus
showing that it is fully determined by the $W_i$, as claimed.
{}For completeness we record the explicit form:
\begin{eqnarray}\label{intrinstorsion}
T^0_{rmn}&=&\frac{1}{20}\big( 2\Omega W_4- \Omega
W_5\big)_{[m}\Omega_{n]r}+\frac{1}{4}(\Omega
W_4)_{r}\Omega_{mn}+3\Omega_{[r}{^k}(W_3)_{mn]k}\nn
&+&\frac{1}{2\cdot 4!}\chi_{rmn}{^{k_1k_2}}(W_1)_{k_1
k_2}-\frac{1}{4!}\Omega_{[m}{^k}(W_2)_{|k\ell_1\ell_2\ell_3|}\chi{^{\ell_1\ell_2\ell_3}}{_{n]r}}
\end{eqnarray}

Now since $\nabla^0\equiv(\nabla+K^0)$ preserves the $SU(5)$
structure, where $\nabla$ is the Levi-Civita connection, it leaves
invariant the spinor $\eta$. Note that the spin connection of
$\nabla^0$ is related, in our conventions, to the spin connection
of the Levi-Civita connection and the contorsion tensor by,
\begin{equation}
\omega^0{_{m}}{^a}{_b}=\omega_m{^a}{_b}+K^{0}{^a}{_{m b}}
\end{equation}
where $K^{0}{^a}{_{m b}}\equiv e^a_r e_b^n K^{0}{^r}{_{mn}}$.
Using this and \p{noname} we see  that the spinor $\eta$ solves
\begin{eqnarray}\label{spincalc}
[\nabla_{m}&+&\frac{1}{160}(\Omega W_5+5\Omega W_4)_m\Omega_{k_1
k_2}\Gamma^{k_1 k_2} -\frac{1}{16}(W_4)_k\Gamma_m{}^k \nn
&+&\frac{1}{8}\Omega_m{^r}(W_3)_{r k_1 k_2}\Gamma^{k_1 k_2}
-\frac{1}{394}\chi_{m k_1 k_2}{}^{n_1n_2}(W_1)_{n_1n_2}
\Gamma^{k_1 k_2}\nn &+&\frac{1}{192}\Omega_m{^r}(W_2)_{r
\ell_1\ell_2\ell_3}\chi^{\ell_1\ell_2\ell_3}{_{k_1
k_2}}\Gamma^{k_1 k_2}]\eta=0
\end{eqnarray}

Let us make two further comments about $SU(5)$ structures that are
not of direct relevance to the derivations in the text. {}Firstly,
having got explicit expressions for the most general $SU(5)$
preserving connection we can easily see which $SU(5)$-structures
admit a connection with totally antisymmetric torsion. {}From
\p{contcomp} we see that this requires that $T^{(3)}$ and
$Q^{2,1}_0$ vanish and also that the structure must satisfy,
\begin{eqnarray}\label{antisymm}
W_2=0\nn 8 W_4+W_5=0
\end{eqnarray}

Secondly lets discuss how an $SU(5)$ structure is affected by a
conformal transformation of the metric. Consider an $SU(5)$
structure $(g,\Omega,\chi)$ and a transformation $g\to
\tilde{g}=e^{2f}g$. The metric $\tilde{g}$ then admits an $SU(5)$
structure $(\tilde{g},\tilde{\Omega},\tilde{\chi})$ which is
related to the original one by,
\begin{eqnarray}
\tilde{\Omega}&=&e^{2f}\Omega\nn \tilde{\chi}&=&e^{5f}\chi
\end{eqnarray}
The components $W_i$ then transform as,
\begin{eqnarray}
\tilde{W}_1&=&e^f W_1\nn \tilde{W}_2&=&e^{3f} W_2\nn
\tilde{W}_3&=&e^{2f}W_3\nn \tilde{W}_4&=&W_4+8df\nn
\tilde{W}_5&=&W_5-40df
\end{eqnarray}
We see that even though the components $W_4$ and $W_5$ transform
non-trivially, the combination $W_5+5 W_4$ is conformally
invariant. A simple corollary is
that a geometry is conformal to a Calabi-Yau five-fold
if and only if it has an $SU(5)$ structure with $W_1=W_2=W_3=0$
and also has $W_4,W_5$ exact and satisfying $W_5=-5W_4$

\section{Some useful identities}
\makeatletter
\renewcommand{\theequation}{D.\arabic{equation}}
\@addtoreset{equation}{section} \makeatother

We record here some of the identities satisfied by irreps of
$SU(5)$ that are useful in deriving the results of table 1 and
other formulae. Let $\Lambda^{(p,q)}$ denote a $(p,q)$ form with a
subscript of $0$ denoting removal of traces, corresponding to an
irreducible representation of $SU(5)$ (see \p{eqn:reps}), then
\bea\label{sufids} \ast
\Lambda^{(3,1)}_{0}&=&-\Lambda^{(3,1)}_{0}\wedge\Omega\nn \ast
\Lambda^{(2,2)}_{0}&=&\Lambda^{(2,2)}_{0}\wedge\Omega\nn \ast
\Lambda^{(2,0)}&=&\frac{1}{3!}\Lambda^{(2,0)}\wedge \Omega^3\nn
\ast
\Lambda^{(1,1)}_{0}&=&-\frac{1}{3!}\Lambda^{(1,1)}_{0}\wedge\Omega^3\nn
\ast(\Lambda^{(2,0)}\wedge\Omega)&=&\frac{1}{2}\Lambda^{(2,0)}\wedge\Omega^2\nn
\ast(\Lambda^{(1,1)}_{0}\wedge\Omega)&=&-\frac{1}{2}
\Lambda^{(1,1)}_{0}\wedge\Omega^2\nn
\Lambda^{(1,1)}_{0}\wedge\Omega^4&=&0\nn
\Lambda^{(3,1)}_{0}\wedge\Omega^2&=&0 \eea

The following identities are useful in deriving \p{finres}:
\begin{eqnarray}\label{spinids}
\Gamma_{a_1....a_4}\e&=&-\frac{i}{2} \theta_{a_1...a_4
b}\Gamma^{b}\e^*\nn \Gamma_{a_1....a_3}\e&=&\frac{i}{2^2 2!
}\theta_{a_1...a_3 b_1 b_2 }\Gamma^{b_1 b_2} \e^* \nn \Gamma_{a_1
a_2}\e&=&\frac{i}{2^3 3!} \theta_{a_1 a_2
b_1...b_3}\Gamma^{b_1...b_3}\e^*\nn \Gamma_{a_1}\e&=&-\frac{i}{2^4
4!}\theta_{a_1 b_1 ...b_4}\Gamma^{b_1...b_4}\e^*\nn
\e&=&-\frac{i}{2^5 5!}\theta_{a_1...a_5}\Gamma^{a_1...a_5}\e^*
\end{eqnarray}

\section{$SU(5)$ structures in eleven dimensions}
\makeatletter
\renewcommand{\theequation}{E.\arabic{equation}}
\@addtoreset{equation}{section} \makeatother

In deriving the form of the most general geometry admitting a
timelike Killing spinor we used the fact that $K$ was a timelike
Killing vector and then worked with the $SU(5)$ structure on the
$D=10$ base manifold, $B$, orthogonal to the orbits of $K$.
Moreover, our analysis determined the type of $SU(5)$ structure on
$B$.

Here we briefly indicate how our analysis also determines the type
of D=11 $SU(5)$ structure specified by $(K,\Omega,\Sigma)$ when
$K^2\neq 0$. {}For example, the fact that $K$ is Killing expresses
the vanishing of some components of the corresponding intrinsic
torsion. One conceptual advantage of discussing the $SU(5)$
structure in $D=11$ is that that the rotation parameter $d\omega$
arises as a component of the structure, while from the ten
dimensional point of view it is just an arbitrary closed two form.

An $SU(5)\subset SO(10,1)$ structure in D=11 can be specified by a
one-form $V$, a two-form $J$ and a five-form $\sigma$ such that
the vector dual to $V$ is timelike, and the forms satisfy
\begin{eqnarray}\label{algebraic}
i_V J=0\nn i_V \sigma=0 \nn J\wedge \sigma=0\nn J\wedge i_V
*\sigma=0\nn \sigma\wedge i_V *\sigma=-2^4\frac{J^5}{5!}
\end{eqnarray}
The one form $V$ allows us to reduce $SO(10,1)\to SO(10)$ and
$(J,\sigma)$ to further reduce $SO(10)\to SU(5)$. We require that
the forms defining the structure have constant norm in the eleven
dimensional metric (and so are related to rescaled versions of
$(K,\Omega,\Sigma)$ as we shall see).

As we discussed in section 3, such structures are classified by
the intrinsic torsion $T^0$ which lives in the space $T^*\otimes
su(5)^\perp$ where $su(5)\oplus su(5)^\perp=so(10,1)$. The adjoint
of $so(10,1)$ decomposes as ${\bf 55}\to {\bf 1}+({\bf
5+\bar{5}})+({\bf 10+\bar{10}})+{\bf 24}$ and so the complement of
$su(5)$ is given by $g^\perp={\bf 1}+({\bf 5+\bar{5}})+({\bf
10+\bar{10}})$. Noting the following decomposition,
\begin{eqnarray}\label{tor}
&&({\bf 1}+{\bf 5+\bar{5}})\otimes  ({\bf 1}+{\bf 5+\bar{5}}+{\bf
10+\bar{10}})\to \nn &&{\bf 1}+{\bf
1'+1''+(5+\bar{5})+(5+\bar{5})'} +{\bf (5+\bar{5})''}+{\bf
(10+\bar{10})+(10+\bar{10})'}+\nn && {\bf
(10+\bar{10})''+(15+\bar{15})} + {\bf 24+24'+
(40+\bar{40})+(45+\bar{45})}
\end{eqnarray}
we see that there are fourteen classes of $SU(5)$ structures in
$D=11$, and we can write
\begin{equation}
T^0\in \bigoplus_{i=1}^{14}{\mathcal W}_i
\end{equation}

The intrinsic torsion in each of the modules ${\mathcal W}_i$ can
be expressed in terms of the exterior derivatives of
$(V,\Omega,\sigma)$. To see this we decompose the exterior
derivatives of the forms defining the structure and see which
representation appear. Taking into account that they are $SU(5)$
invariant, we find
\begin{eqnarray}\label{dtor}
dV &\to& {\bf 1+24+(10+\bar{10})+(5+\bar{5})}\nn dJ &\to& {\bf
(5+\bar{5})+(10+\bar{10})+(45+\bar{45})+1+(10+\bar{10})'+24}\nn
d\sigma &\to &  {\bf
(5+\bar{5})+(10+\bar{10})+(40+\bar{40})+1+1'+(10+\bar{10})'}\nn
&+&{\bf (15+\bar{15})}
\end{eqnarray}
In comparing these to the irreps appearing in the intrinsic
torsion, there appears to be a mismatch since four ${\bf 1}$'s and
five ${\bf 10+\bar{10}}$'s appear in \p{dtor} while in \p{tor} we
have only three  ${\bf 1}$'s and three ${\bf 10+\bar{10}}$'s.
However this is not so since we can relate the ${\bf 1}$ of $dJ$
to the ${\bf 1+1'}$ of $d\sigma$ by the last condition in
\p{algebraic}, while the ${\bf (10+\bar{10})}$ of $dJ$ and
$d\sigma$ are related by $J\wedge\sigma=0$ and similarly for the
${\bf (10+\bar{10})'}$.

As mentioned, the forms used above to define an $SU(5)$ structure
in D=11 are not the ones constructed from the Killing spinors.
They are related to those, in regions where $K^2=-\Delta^2\neq 0$, by
\begin{eqnarray}
K&=&\Delta V\nn \Omega&=&\Delta J\nn K\wedge \Sigma&=&\Delta^2
V\wedge \sigma
\end{eqnarray}
Given that we know expressions for $dK,d\Omega, d\chi$ we can
obtain those for $dV, dJ, d\sigma$ and hence precisely determine
the restrictions placed on the D=11 ${\mathcal W}_i$. In other
words our analysis does indeed determine the D=11 $SU(5)$
structure as claimed.


\end{document}